\documentclass[twocolumn,english,aps,prl,floatfix,preprintnumbers,showpacs,amsfonts,amssymb,superscriptaddress]{revtex4}
\usepackage{ae,aecompl}
\usepackage[T1]{fontenc}
\usepackage[latin1]{inputenc}
\usepackage{array}
\usepackage{multirow}
\usepackage{amsmath}
\usepackage{amssymb}
\usepackage{graphicx}

\usepackage{color}
\usepackage[bookmarks=true,colorlinks,linkcolor=blue,urlcolor=blue,citecolor=blue]{hyperref}

\newcommand{\be}{\begin{equation}}
\newcommand{\ee}{\end{equation}}
\newcommand{\bea}{\begin{eqnarray}}
\newcommand{\eea}{\end{eqnarray}}
\newcommand{\mc}{\mathcal}

\DeclareMathOperator{\sgn}{sgn}
\DeclareMathOperator{\im}{Im}

\makeatletter

\providecommand{\tabularnewline}{\\}
\@ifundefined{textcolor}{}
{%
 \definecolor{BLACK}{gray}{0}
 \definecolor{WHITE}{gray}{1}
 \definecolor{RED}{rgb}{1,0,0}
 \definecolor{GREEN}{rgb}{0,1,0}
 \definecolor{BLUE}{rgb}{0,0,1}
 \definecolor{CYAN}{cmyk}{1,0,0,0}
 \definecolor{MAGENTA}{cmyk}{0,1,0,0}
 \definecolor{YELLOW}{cmyk}{0,0,1,0}
}

\usepackage{bm}

\makeatother

\usepackage{babel}
\begin{document}

%\title{Boundary exponents for nonlinear Luttinger liquids}
\title{Boundary versus bulk behavior of time-dependent correlation functions in one-dimensional quantum systems}

\author{I. S. Eli\"ens}
\affiliation{Institute for Theoretical Physics, Institute of Physics, University of Amsterdam, Science Park 904, 1098 XH Amsterdam, The Netherlands }
 \email{i.s.eliens@uva.nl}
 \author{F. B. Ramos}
 \affiliation{Universidade Federal de Uberl\^andia, Instituto de F\'isica, C.P. 593, 38400-902 Uberl\^andia, MG, Brazil}
 \author{J. C. Xavier}
 \affiliation{Universidade Federal de Uberl\^andia, Instituto de F\'isica, C.P. 593, 38400-902 Uberl\^andia, MG, Brazil}
\author{R. G. Pereira}
\affiliation{Instituto de F\'{i}sica de S\~ao Carlos, Universidade de S\~ao Paulo, C.P. 369, S\~ao Carlos, SP, 13560-970, Brazil}

\date{\today} 

\begin{abstract}
We study the influence of reflective boundaries on time-dependent responses of one-dimensional quantum fluids at zero temperature beyond the low-energy approximation. Our analysis is based on an extension of effective mobile impurity models for  nonlinear Luttinger liquids to the case of open boundary conditions. For integrable models, we show that  boundary autocorrelations oscillate as a function of time with the same frequency as  the corresponding bulk autocorrelations. This frequency can be identified as the band edge of elementary excitations. The amplitude of the oscillations decays as a power law with distinct  exponents  at the boundary and in the bulk, but  boundary and bulk exponents are determined by the same coupling constant in the mobile impurity model.  For nonintegrable models, we argue that the power-law decay of the oscillations is generic for autocorrelations in the bulk, but turns into an exponential  decay at the boundary. Moreover, there is in general a nonuniversal   shift of the boundary frequency in comparison with the band edge  of bulk excitations.  The predictions of our effective field theory are compared with numerical results obtained by
time-dependent density matrix renormalization group (tDMRG) for both
integrable and nonintegrable critical spin-$S$ chains with $S=1/2$, $1$  and $3/2$. 

\end{abstract}

\pacs{71.10.Pm,75.10.Pq} 
\maketitle

\section{Introduction}
Striking  properties in many-body quantum systems often emerge  from the
interplay between interactions and a constrained geometry.
In a Fermi gas confined to a single spatial dimension, for example,
interactions lead to
dramatically different spectral properties as compared to its higher
dimensional counterparts described by Fermi liquid theory
\cite{1967_Theumann_JMP_8,1974_Luther_PRB_9,1992_Meden_PRB_46,1993_Voit_JPCM_5}.

The low-energy limit of one-dimensional (1D) Fermi gases is conventionally treated within the
Luttinger liquid (LL) framework \cite{GiamarchiBOOK}. Indispensable in this
respect is the
exactly solvable  Tomonaga-Luttinger (TL) model
\cite{1950_Tomonaga_PTP_5,1963_Luttinger_JMP_4}, which allows a
nonperturbative treatment of  interactions at the cost of an  artificially linearized  dispersion relation for the constituent fermions.  
Using the technique of bosonization,  
the model is solved in terms of bosonic collective modes
corresponding to quantized waves of density. 

Static correlations and many thermodynamic properties are
  captured remarkably well by the Luttinger liquid approach.  For many
  dynamic effects, however, it is clear that band curvature 
  needs to be taken into account. For example, the relaxation of the
  bosonic sound modes, or the related width of the dynamical structure
  factor (DSF), are not captured by Luttinger liquid theory, which
  predicts a delta function peak for the DSF.  Attempts to treat the
  DSF broadening in the bosonized theory, in which the dispersion
  curvature translates to interactions between the modes diagonalizing
  the TL model, are hindered by on-shell divergences in
  the perturbative expansion. Certain aspects of the DSF broadening
  can nevertheless be captured in the bosonic basis
  \cite{1998_Samokhin_JPC_10,2007_Aristov_PRB_76,
    2007_Teber_PRB_76,2006_Teber_EPJB_52,2014_Price_PRB_90}.  An
  alternative approach uses a reformulation of the TL model including
  a quadratic correction to the dispersion in terms of fermionic
  quasiparticles.  In the low-energy limit, these
  turn out to be weakly interacting \cite{2005_Rozhkov_EPJB_47,
    2007_Khodas_PRB_76,2009_Imambekov_SCIENCE_323} restoring some of
  the elements of Fermi liquid theory in one dimension. 
  At high energies, insight into dynamic
  response functions such as the DSF and the spectral function, and in
  particular into the characteristic threshold singularities, can be
  obtained by mapping the problem to a mobile impurity Hamiltonian. This
  approach hinges on the observation that the thresholds correspond to
  configurations of a high energy hole or particle which can
  effectively be considered as separated from the low energy subband, and
  that the threshold singularities emerge from the scattering
 of the modes at the Fermi level on this impurity mode.  This identifies
  the anomalous correlation structure of 1D gases as an example of
  Anderson's orthogonality catastrophe \cite{1967_Anderson_PRL_18} and
  links it to the physics of the x-ray edge singularity
  \cite{1969_Schotte_PR_182}.  Many new results on dynamic
  correlations, in general
  and for specific models, have been
  obtained this way
  \cite{2003_Carmelo_PRB_68,2006_Pereira_PRL_96,2006_Pustilnik_PRL_96,
    2007_Khodas_PRL_99,2007_Khodas_PRB_76,2006_Pustilnik_PRL_96,
    2008_Pereira_PRL_100,2008_Cheianov_PRL_100,2009_Pereira_PRB_79,
    2009_Imambekov_PRL_102,2010_Essler_PRB_81,2010_Schmidt_PRB_82,
    2010_Schmidt_PRL_104,2012_Pereira_PRB_85}.
  This bears relevance to e.g.  Coulomb drag experiments \cite{2002_Debray_SST_17,2002_Yamamoto_PE_12,2003_Pustilnik_PRL_91,2007_Aristov_PRB_76,2011_Laroche_NATNANO_6,2010_Pereira_PRB_82}
  as well as relaxation and transport
  \cite{2013_Matveev_PRL_111,2014_Protopopov_PRB_90,
    2014_Protopopov_PRB_89,2013_Lin_PRL_110,2010_Barak_NATPHYS_6}. Dispersion
  nonlinearity also greatly influences the propagation of a density
  bump or dip, which would retain its shape when time-evolved under
  the linear theory but relaxes by emitting shock waves in the
  nonlinear theory \cite{2012_Bettelheim_PRL_109,
    2006_Bettelheim_PRL_97,2013_Protopopov_PRB_87}.
  Closer to the
  present work is the late-time dependence of correlations
  \cite{2012_Pereira_IJMPB_26,2015_Karrasch_NJP_17,2014_Seabra_PRB_90}
  which are related to the singularities in the
  frequency domain. Collectively, the extensions of LL theory that
  include band curvature effects may be called nonlinear
  Luttinger liquid (nLL) theory, but we will mainly be concerned
  with the mobile impurity approach to correlations (see Ref.~\onlinecite{2012_Imambekov_RMP_84} for  further details).

Motivated by these theoretical advances, we study the effect of reflective
boundaries on a 1D gas beyond the low-energy regime. 
Our work is also inspired by    studies of ``boundary critical phenomena'' \cite{1989_Cardy_NPB_324,1994_Affleck_JPA_27,1996_Fendley_JSP_85} within the LL
framework that   have unveiled  remarkable effects, e.g.,   in the conductance  of quantum wires \cite{1992_Kane_PRB_46, 1994_Wong_NPB_417,1995_Fabrizio_PRB_51},  screening of magnetic  impurities \cite{1992_Eggert_PRB_46},  
Friedel oscillations in charge and spin densities
\cite{1995_Egger_PRL_75,1996_Leclair_PRB_54,2000_Rommer_PRB_62}, and oscillations  in the entanglement entropy \cite{2006_Laflorencie_PRL_96,2013_Taddia_PRB_88}.

We  focus on response functions which can be locally addressed---such
as the local density of states (LDOS) and autocorrelation
functions---as these are expected to show the clearest bulk
versus boundary contrast. % and have physically transparent interpretation.
Many studies have addressed the LDOS for LLs with a boundary
\cite{2000_Schoenhammer_PRB_61,2000_Meden_EPJB_16,2002_Schollwoeck_PTPS_145,
2002_Meden_PRB_65,2004_Andergassen_PRB_70,2006_Andergassen_PRB_73,2008_Schneider_PRL_101,2013_Schuricht_JPCM_25,2013_Soeffing_EPL_101,2013_Jeckelmann_JPCM_25}.
LL theory predicts a characteristic power-law suppression (for repulsive interactions)
of the LDOS at the Fermi level with  different bulk
and boundary exponents which are nontrivially but universally related 
\cite{1996_Eggert_PRL_76,1997_Mattsson_PRB_56}. This has been
verified using different techniques
\cite{1996_Wang_PRB_54,2000_Schoenhammer_PRB_61,2000_Meden_EPJB_16,2013_Schuricht_JPCM_25}
and is used as a consistency check in the experimental  identification of  LL physics \cite{1999_Bockrath_NATURE_397,2011_Blumenstein_NATPHYS_7}.

Away from the Fermi level, no universal results are known. This
pertains  both to   general
statements on the restricted energy range where the power-law scaling is valid 
%which has been found to be exponentially small for the Hubbard model
\cite{2000_Meden_EPJB_16,2013_Schuricht_JPCM_25} 
and  to 
 details of the line shape at higher energies. Here, we deal with the
 latter and argue that the nonanalyticities of, e.g., the LDOS away from zero energy   can be
understood in the framework of nLL theory for
systems with open   and periodic boundary
conditions alike. The main application of our theory is in describing   the power-law decay of  autocorrelation functions in real  time. We show that bulk and boundary exponents  are governed by the same parameters  in the mobile impurity model and obey   relations that  depend only on the Luttinger parameter. 
These relations provide a quantitative test of the nLL theory.  We perform this test by analyzing  time-dependent density matrix renormalization group (tDMRG) \cite{2004_White_PRL_93,2011_Schollwock_AnnPhys_326} results for   spin autocorrelations  of critical spin   chains. The statement about boundary exponents applies to integrable models in which  the nonanalytic behavior at finite energies is not susceptible to broadening due to three-body scattering processes \cite{2007_Khodas_PRB_76,2009_Pereira_PRB_79}. The effects of integrability breaking are also investigated, both numerically and from the perspective of the mobile impurity model. We find that for nonintegrable models the finite-energy singularities in boundary autocorrelations  are broadened by decay processes associated  with boundary operators in the mobile impurity model. As a result, the boundary autocorrelation decays exponentially in time in the  nonintegrable case.

The paper is organized as follows. In Section \ref{sec:GF}, we discuss the LDOS  for spinless fermions as a first  example of how dynamical correlations in the vicinity of an open boundary differ from the result in the bulk. In Section \ref{sec:impuritymodel}, we present the mobile impurity model used to calculate  the exponents in the LDOS near the boundary. In Section \ref{sec:other}, we generalize our approach to predict relations between   bulk and boundary exponents of other dynamical  correlation functions, including the case of spinful fermions. Section \ref{sec:integrability} addresses the question  whether finite-energy singularities exist in nonitegrable models. Our numerical results for the time decay of spin autocorrelation functions are presented in Section \ref{sec:tDMRG}. Finally,  we offer some concluding remarks in Section \ref{sec:conclusion}.

%%%%%%%%%%%%%%%%%%%%%%%%%%%%%%%%%%%%%%%%%%%%%%%%%%%%%%%%%%%%%%%%%%%%%%%%%%%%%%%%%%%%

\section{Green's function for spinless fermions\label{sec:GF}}

We are interested in 1D systems  on a half-line, where we impose the boundary condition  that all physical operators vanish at $x=0$. Let us first discuss the case of spinless fermions on a lattice.  We define the (non-time-ordered) Green's function at  position $x$ as
\begin{equation}
\label{eq:5}
G(t,x) = \langle\{ \Psi(x,t),\Psi^{\dag}(x,0)\}\rangle, 
\end{equation}
where $\Psi(x)$ annihilates a spinless fermion at position $x$ and the time evolution $\Psi(x,t)=e^{iHt}\Psi(x)e^{-iHt}$ is governed by a local Hamiltonian $H$.  The brackets $\langle\ldots \rangle$ denote the expectation value in the ground state of $H$. The Fourier transform to the frequency domain yields the LDOS \be
\rho(\omega,x)=\frac1{2\pi}\int_{-\infty}^{\infty}dt\, e^{i\omega t} G(t,x).
\ee

The boundary case   corresponds to the result  for $x=a$, where $a$ is the lattice spacing for lattice models or the short-distance cutoff    for continuum models. We refer to the bulk case of $G(t,x)$ as the regime $x\gg a$ and $vt<x$, where $v$ is the velocity that sets the light cone for propagation of correlations in the many-body system \cite{1972_Lieb_CommMathPhys_3}. The latter  condition allows one to neglect the effects of reflection at the boundary, and is routinely employed in numerical simulations aimed at capturing the long-time behavior   in the thermodynamic limit \cite{2008_Pereira_PRL_100,2009_Pereira_PRB_79,2014_Karrasch_PRB_89,2014_Seabra_PRB_90}. 

As our point of departure, consider the free fermion model \bea
H_0&=&-\frac{1}{2}\sum_{x\geq1}[\Psi^\dagger(x){\Psi^{\phantom\dagger}(x+1)}+\text{h.c.}]\nonumber\\
&=&\sum_{k}\varepsilon_k\Psi^\dagger_k \Psi^{\phantom\dagger}_k,\label{H0}\eea
 where $\varepsilon_k=-\cos k$, with $k\in(0,\pi)$, is the free fermion dispersion  and we set $a=1$. The single-particle eigenstates of $H_0$ are created by \be
 \Psi^\dagger_k=\sqrt{\frac2\pi}\sum_{x\geq1}\sin(kx)\Psi^\dagger(x).\ee 
 We focus on the case of half filling, in which the ground state is constructed by occupying  all states with $0<k<\pi/2$. In this case  particle-hole symmetry rules out Friedel oscillations \cite{2000_Rommer_PRB_62} and the average density is homogeneous, $\langle \Psi^\dagger(x)\Psi(x)\rangle=1/2$. The Green's function is given exactly by \be
G_0(t,x)=\frac{4}{\pi}\int_0^{\pi/2}dk\, \sin^2(kx)\cos(\varepsilon_k t ),\label{G0free}
\ee 
and the LDOS is
\be
\rho_0(\omega,x)=\frac{2\sin^2[x\arccos(\omega/\epsilon_0)]}{\pi\sqrt{\epsilon_0^2-\omega^2}}\theta(\epsilon_0-|\omega|),\label{rho0w}
\ee
where $\epsilon_0\equiv |\varepsilon_{k=0}|=1$.

\begin{figure}
\begin{center}
\includegraphics*[width=\columnwidth]{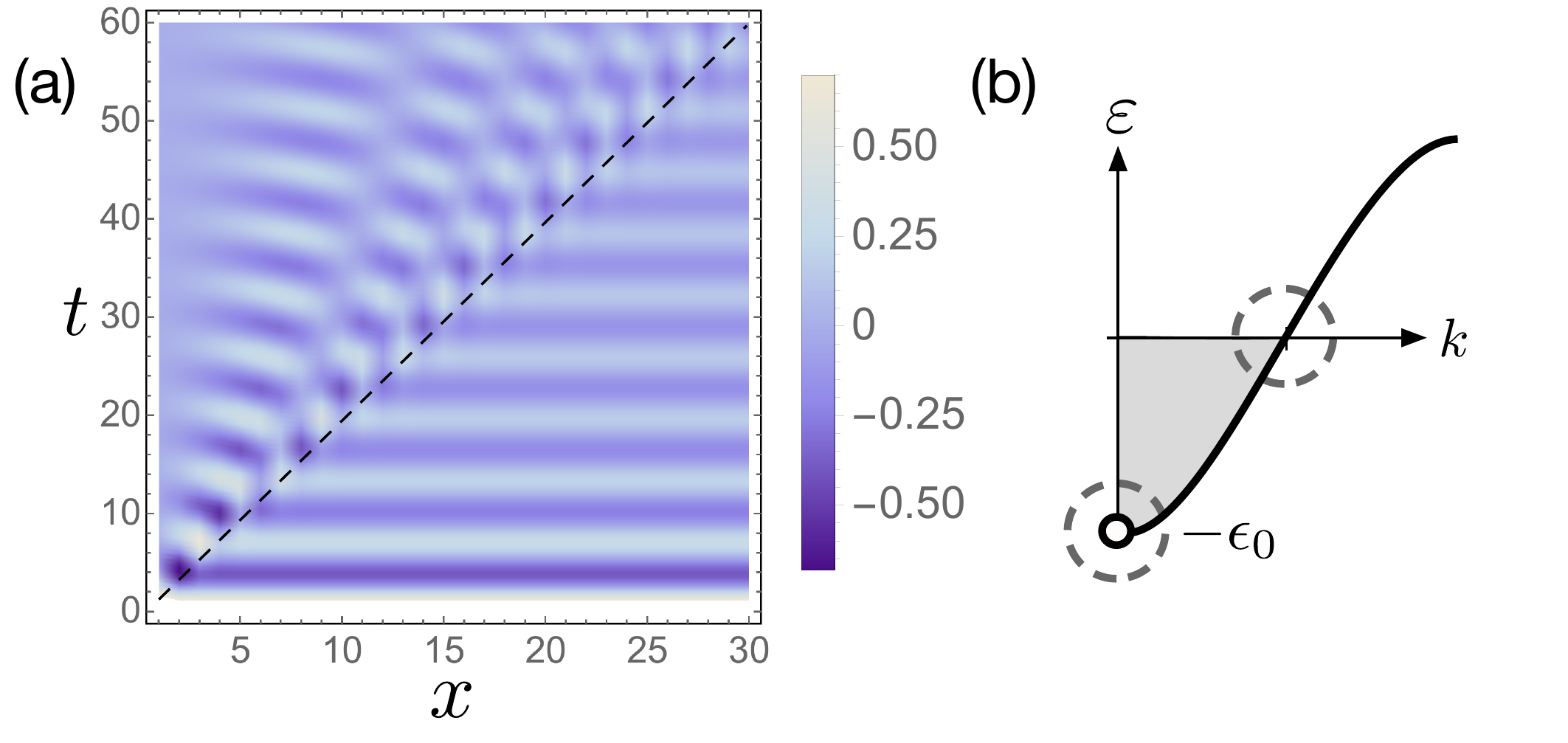}
\end{center}
\caption{(Color online) 
(a) Green's function $G_0(t,x)$  for free fermions in a
semi-infinite chain at half-filling [Eq. (\ref{G0free})], where $x$ is
the distance from the boundary. The dashed line represents the
reflection time $T_{\text{refl}}(x)=2x/v$ with $v=1$. (b) The deep
hole configuration responsible for the oscillations at $x=0$ related
to the singularities of the LDOS (Fig. \ref{fig:regimes}).  There is an
equivalent high-energy particle configuration, not depicted. The
dashed circles indicate the projection onto low-energy and impurity
subbands important once interactions are taken into account.
}\label{fig:cone}
\end{figure}

The result for $G_0(t,x)$   is depicted in Fig. \ref{fig:cone} (a).  First we note that, for any fixed position $x$, there is a clear change of behavior  at the time scale $t\sim T_{\text{refl}}(x)=2x/v$ (where  $v=1$   for free fermions).  This corresponds to the time for the light cone centered at $x$ to reflect at the boundary and return to $x$. For $t< T_{\text{refl}}(x)$, $G_0(t,x)$  is independent of $x$ ({\it i.e.} translationally invariant for fixed $t$ and  $x>vt/2$) and the result is representative of the bulk autocorrelation. The arrival of the  boundary-reflected correlations makes $G_0(t,x)$ deviate from the bulk case and become $x$-dependent for $t>T_{\text{refl}}(x)$.  After we take the Fourier transform to the frequency domain, the reflection time scale   implies that   the LDOS  in Eq. (\ref{rho0w})  oscillates with  period $\Delta \omega(x)\sim 2\pi/T_{\text{refl}}(x)=\pi v/x$.  In the bulk case, the rapid oscillations in the frequency dependence of $\rho_0(\omega,x\gg 1)$  are averaged out by any finite frequency resolution \cite{2013_Jeckelmann_JPCM_25}.  In  numerical simulations of time evolution  in the bulk, the usual procedure is to stop the simulation at   $t<x/v$ (or before in case  the maximum time is limited by various sources of  error \cite{2004_White_PRL_93,2011_Schollwock_AnnPhys_326}). This avoids the reflection at the boundary but at the same time sets the finite frequency resolution. 
 
\begin{figure}
\begin{center}
\includegraphics*[width=.85\columnwidth]{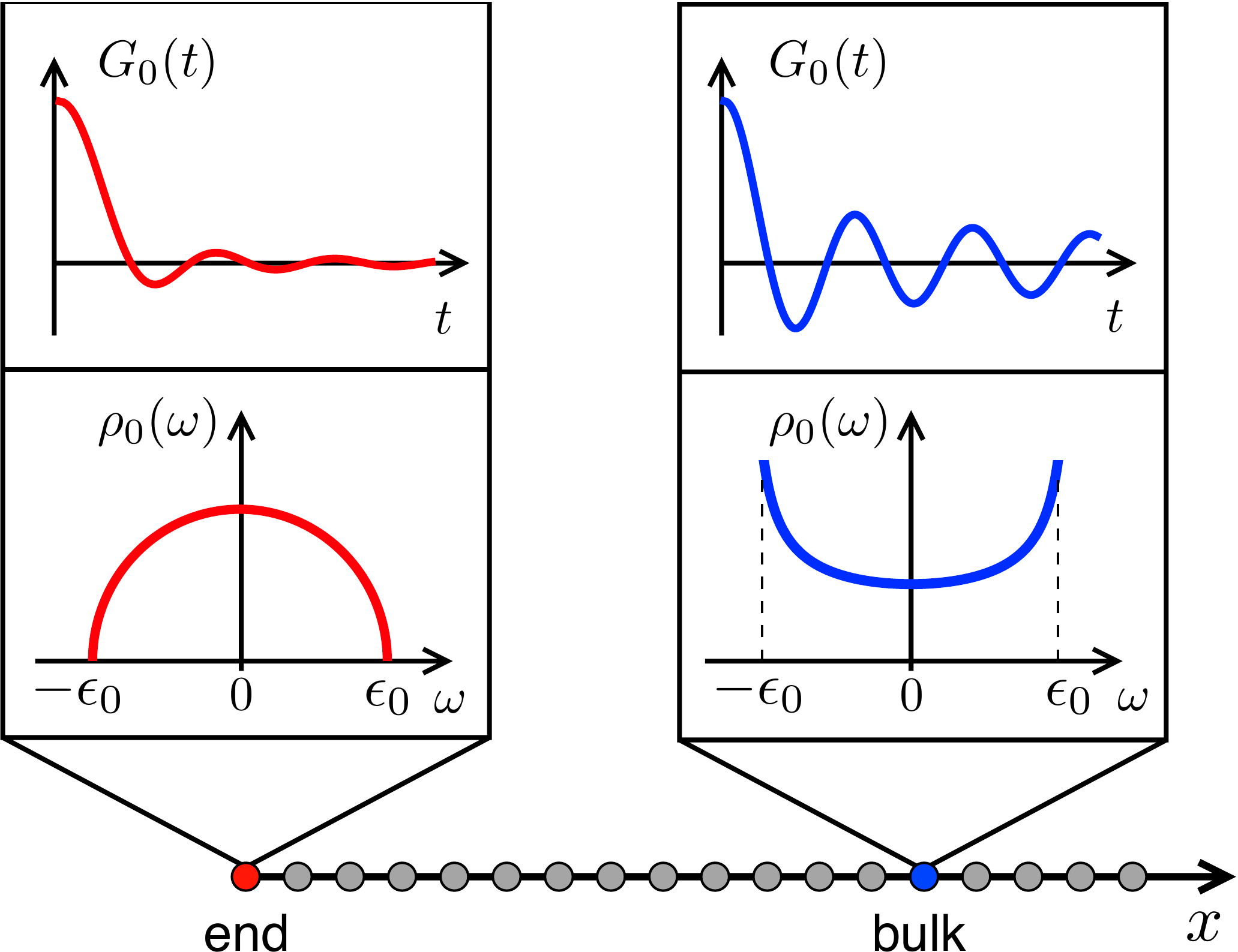}
\end{center}
\caption{ (Color online)
Noninteracting Green's function $G_0(t,x)$ and LDOS $\rho_0(\omega,x)$. The curves on the left correspond to the  chain end ($x=1$), and the curves  on the right to a site in the bulk ($x\gg1$). 
}\label{fig:regimes}
\end{figure}

Let us now discuss the time dependence of the Green's function at the boundary ($x=1$) versus in the bulk ($x\gg 1$, $vt<x$).  In both cases  (see Fig. \ref{fig:regimes}) the Green's function shows oscillations in the long-time decay which are not predicted by the usual low-energy approximation of linearizing the dispersion about $k_F=\pi/2$ \cite{GiamarchiBOOK}. 
The explanation for the real-time oscillations is the same for  open
or periodic boundary conditions; for the case of periodic boundary
conditions, see the reviews in
Refs. \cite{2012_Imambekov_RMP_84,2012_Pereira_IJMPB_26}. The
oscillations stem from a saddle point contribution to the integral in
Eq. (\ref{G0free}) with  $k\approx 0$ [in the hole term of $G_0(t,x)$]
or $k\approx \pi$ (in the particle term). This contribution is
associated with an excitation with  energy $\epsilon_0$, the maximum
energy of a single-hole or single-particle excitation [see
Fig. \ref{fig:cone} (b)]. We call this energy the band edge of the free fermion dispersion. The propagator of   the band edge mode decays more slowly in time due to its  vanishing group velocity. The importance of  this finite-energy contribution is manifested  in the LDOS as a power-law singularity at $\omega= \pm \epsilon_0$ (see Fig. \ref{fig:regimes}). Notice the clear difference between the bulk and the boundary case: while in the bulk the  LDOS has a van Hove singularity at the band edge, $\rho_0(\omega,x\gg1)\sim |\omega \pm\epsilon_0|^{-1/2}$, at the boundary one finds  a square-root cusp $\rho_0(\omega,x=1)\sim |\omega\pm\epsilon_0|^{1/2}$.

One of the main achievements of the nLL theory is to incorporate the contributions of finite-energy excitations in dynamical correlation functions for interacting 1D systems with band curvature \cite{2012_Imambekov_RMP_84,2012_Pereira_IJMPB_26}. Our purpose here is to generalize this approach  to describe the dynamics in the vicinity of a boundary. For concreteness, we  consider the  model\be
H=H_0+V\sum_{x\geq1}n(x)n(x+1),\label{Hinteracfermion}
\ee
where $n(x)\equiv \Psi^\dagger(x)\Psi(x)$ is the density operator and we focus on the repulsive regime $V>0$. Importantly, the model in Eq. (\ref{Hinteracfermion}) is integrable and exactly solvable by Bethe ansatz \cite{KorepinBOOK}.  This guarantees that the band edge of elementary excitations is still well defined in the interacting case. We postpone a detailed  discussion about integrability-breaking effects to Section \ref{sec:integrability}.

Before outlining the derivation of the results for  the interacting model (see Section \ref{sec:impuritymodel}), we summarize some known results together with  our findings for the Green's function and LDOS. The calculation within the LL framework leads
to the well-known predictions \cite{1992_Kane_PRB_46,1996_Eggert_PRL_76,1997_Mattsson_PRB_56}
\bea
\label{eq:31}
 G_{\text{LL}}(t,x) &\sim &1/t^{\alpha + 1},\\
\rho(\omega\approx 0,x) &\sim &|\omega|^{\alpha},\label{rhoLL}
\eea
where the exponent $\alpha$ is different   for $x$ in the bulk  than at the boundary (subscript ``end''): $ \alpha_{\text{bulk}} = (K + K^{-1})/2-1$ and $
 \alpha_{\text{end}} = K^{-1} -1$,
where $K$ is the Luttinger parameter ($K=1$ for free fermions and $K<1$ for repulsive interactions). As mentioned above, the real-time oscillations are not predicted by LL theory. 
It is known that taking into account the finite-energy contributions
within the nLL theory leads to the following contributions
from the band-edge excitation  in the bulk:
\bea
\label{eq:7}
 G_{\text{osc}}(t,x\gg 1)& \sim &e^{\pm i\epsilon t}/t^{\bar\alpha_{\text{bulk}}+1},\\ 
\rho(\omega\approx \pm\epsilon,x\gg1) &\sim& |\omega\mp\epsilon|^{\bar\alpha_\text{bulk}},
\eea
where $\epsilon$ is the renormalized band edge in the interacting
system and the bulk exponent for the oscillating contribution is
\bea
\label{eq:10a}
\bar\alpha_{\text{bulk}} &=&-1/2 + {\gamma^2}/(2\pi^2 K),
\eea 
with $\gamma$ the phase shift of low-energy modes due to scattering off the high-energy hole [for free fermions, ${\gamma=0}$; the phase shift for the  interacting model in Eq. (\ref{Hinteracfermion}) will  be specified in Section \ref{sec:impuritymodel}].

Our new result   is  that the oscillating contribution at the boundary is given by 
\bea
\label{eq:7prime}
 G_{\text{osc}}(t,x= 1)& \sim &e^{\pm i\epsilon t}/t^{\bar\alpha_{\text{end}}+1},\\ 
\rho(\omega\approx \pm\epsilon,x=1) &\sim& |\omega\mp\epsilon|^{\bar\alpha_\text{end}},
\eea
with the same band-edge frequency $\epsilon$ as in the bulk, but with  a different exponent
\bea
\label{eq:10b}
\bar\alpha_{\text{end}} &=& 1/2 + {\gamma^2}/(\pi^2 K).
\eea
When the band-edge mode is the dominant finite-energy contribution to the Green's function, the asymptotic long-time decay of $G(t,x)$ is well described by a linear combination  of the Luttinger liquid term in Eq. (\ref{eq:31}) and the oscillating term in Eq. (\ref{eq:7}) or Eq. (\ref{eq:7prime}).

There are two noteworthy modifications in going from the bulk to the boundary: ({\it i}) an extra factor of $1/t$ in the decay of $G_{\text{ osc}}(t,x)$; ({\it ii})  the doubling of the $\mc O(\gamma^2)$ orthogonality catastrophe correction  to the exponent \cite{1967_Anderson_PRL_18,GiamarchiBOOK}. Both are recurrent in the exponents that will be discussed in Section \ref{sec:other}. Furthermore, while both  exponents vary with  interactions, Eqs. \eqref{eq:10a} and \eqref{eq:10b} imply the relation\be
\bar\alpha_{\text{end}}-2\bar\alpha_{\text{bulk}}=3/2,
 \ee
which is independent of the nonuniversal phase shift $\gamma$.

%%%%%%%%%%%%%%%%%%%%%%%%%%%%%%%%%%%%%%%%%%%%%%%%%%%%%%%%%%%%%%%%%%%%%%%%%%%%%%%%%%%%

\section{Mobile impurity model with open boundary\label{sec:impuritymodel}}

To derive the results above, we %take the continuum limit and 
use the mode expansion that includes band-edge excitations \be
\label{eq:12}
\Psi(x)\sim e^{ik_Fx}\psi_R(x)+e^{-ik_Fx}\psi_L(x)+d^\dagger(x),
\ee
where $\psi_{R,L}$ denote the low-energy modes,
$d^{\dag}$ creates a hole in the bottom of the band ($k\approx 0$), and all fields on
the right-hand side are slowly varying on the scale of the short-distance cutoff $a$. 

A crucial assumption implicit in  Eq. (\ref{eq:12}) is that we  identify the excitations  governing  the long-time decay in the interacting model as being  ``adiabatically connected'' with those in the noninteracting case, in the sense that they carry the same quantum numbers and their dispersion relations vary smoothly as a function of interaction strength. This condition can be verified explicitly for integrable models, where one computes exact dispersion relations for the elementary excitations.  We should also note that for lattice models such as Eq. (\ref{Hinteracfermion}) the mode expansion must   include a high-energy particle at the top of the band, with $k\approx \pi$ \cite{2008_Pereira_PRL_100}. In the particle-hole symmetric case the latter yields a contribution  equivalent to that of the  deep hole with $k\approx 0$, and we get the particle contribution in the LDOS simply by taking $\omega\to -\omega$ in the result for the hole contribution. More generally, the high-energy spectrum of the interacting model may  include other particles and bound states, which can also be incorporated in the  mobile impurity model \cite{2009_Pereira_PRB_79}; we shall address  this question in Section \ref{sec:boundstates}.

In Eq. (\ref{eq:12})
we deliberately write the right and left movers separately, even though they are coupled by the boundary conditions \cite{1992_Eggert_PRB_46,1995_Fabrizio_PRB_51}. The  condition $\Psi(0)=0$ is satisfied if we impose\be
\psi_L(0)=-\psi_R(0),\quad d(0)=0.
\ee 
These relations can be checked straightforwardly  in the noninteracting case using  the single-particle modes $\Psi_k$. The boundary condition on $d(x)$ means that for any boundary operator that involves the high-energy mode we must take $d(a)\sim a\partial_xd(0)$. 

We bosonize the low-energy modes with the conventions \bea
\psi_{R,L}&\sim& e^{-i\sqrt{2\pi}\phi_{R,L}},\\
\psi^\dagger_{R,L}\psi^{\phantom\dagger}_{R,L}&\sim& \mp \frac{1}{\sqrt{2\pi}}\partial_x\phi_{R,L},\eea
 where $\phi_{R,L}(x)$ are chiral bosonic fields that obey $[\partial_x\phi_{R,L}(x),\phi_{R,L}(x^\prime)]={\pm i\delta(x-x^\prime)}$. A convenient way to treat the boundary conditions for the low-energy modes is to use the folding trick \cite{1994_Wong_NPB_417,1995_Fabrizio_PRB_51}: we include negative coordinates $x<0$ and identify \be
\psi_L(x)\equiv -\psi_R(-x).\label{foldingpsi}\ee For the bosonic fields, we use \be
\phi_L(x)\equiv\phi_R(-x)+\sqrt{\pi/2}.\ee 

The effective Hamiltonian that describes the interaction between the band-edge mode and the low-energy modes is the mobile impurity model\bea
\hspace*{-.5cm} H_{\text{MIM}}&=&\int_{-\infty}^{\infty} dx\, \frac{v}2(\partial_x\varphi)^2+\int_0^\infty dx\, d^\dagger\left(\epsilon+\frac{\partial_x^2}{2M}\right)d\nonumber\\
&&+\frac{v\gamma}{\sqrt{2\pi K}}\int_0^\infty dx\,d^\dagger d[\partial_x\varphi(x)+\partial_x\varphi(-x)].\
 \label{eq:Hdensity}
\eea
Here $\varphi(x)$ is the chiral boson  that diagonalizes the Luttinger model  on the unfolded line\be
\varphi(x)=\frac{K^{-\frac12}+K^{\frac12}}2\phi_R(x)+\frac{K^{-\frac12}-K^{\frac12}}2\phi_R(-x),
\ee
which obeys $[\partial_x\varphi(x),\varphi(x^\prime)]=i\sgn(x)\delta(x-x^\prime)$.
The parameters $\epsilon$, $-M$ and $\gamma$ are nonuniversal
properties of the hole with $k=0$ (which is treated as a mobile
impurity): its finite energy cost,  effective mass and dimensionless
coupling to the low-energy modes, respectively. Note that the
  linear term in the dispersion vanishes for the band-edge mode, which
is why we have to take into account the effective mass [see
Fig. \ref{fig:cone} (b)]. In models solvable by Bethe ansatz, $\epsilon$ and $M$ are determined by the exact dispersion of single-hole excitations. The coupling  $\gamma$ can be obtained from the so-called shift function \cite{2008_Cheianov_PRL_100,2008_Imambekov_PRL_100} and the finite size spectrum \cite{2009_Pereira_PRB_79} for  periodic boundary conditions. In
  Galilean-invariant systems, we can relate $\gamma$   to  the exact spectrum by using phenomenological relations \cite{2009_Imambekov_PRL_102}.   

The Hamiltonian in Eq.  (\ref{eq:Hdensity}) contains only marginal operators. It  can be obtained from the mobile impurity model   in the bulk \cite{2009_Imambekov_SCIENCE_323} by applying the folding trick. 
Remarkably, all  boundary operators that perturb this Hamiltonian  and couple the $d$ field to the bosonic modes are highly irrelevant, as they necessarily involve the derivative $\partial_xd(0)$ (which by itself has scaling dimension 3/2). For the moment we neglect the effect of all formally irrelevant boundary operators, but return to this point in Section \ref{sec:integrability}. 

Like in the bulk case, we can decouple the impurity mode by the unitary transformation \begin{equation}
\label{eq:16}
  U = \exp\left\{ i\frac{\gamma}{\sqrt{2\pi K}}\int_{0}^{\infty}dx\, [\varphi(x) +  \varphi(-x)]d^{\dag}d  \right\}.
\end{equation} 
The fields transform as \bea
\tilde \varphi(x)&=&U \varphi(x)U^\dagger=\varphi(x)+ \frac{\gamma}{2\sqrt{2\pi K}}F_d(x),\label{shiftphi}\\
 \tilde d(x)&=&U d(x)U^\dagger= d(x)e^{ -i \frac{\gamma}{\sqrt{2\pi K}}[\varphi(x)  + \varphi(-x)]},\eea
 where \be
F_d(x)=\int_0^\infty dy\, [\text{sgn}(x-y)+\text{sgn}(x+y)]d^\dagger(y)d(y).
 \ee
Eq. (\ref{shiftphi}) implies\be
\partial_x\tilde\varphi(x)=\partial_x\varphi(x)+\frac{\gamma}{\sqrt{2\pi K}}d^\dagger(x)d(x).
\ee
The Hamiltonian becomes noninteracting when written in terms of the transformed fields\be
H_{\text{MIM}}=\int_{-\infty}^{\infty} dx\, \frac{v}2(\partial_x\tilde\varphi)^2+\int_0^\infty dx\, \tilde d^\dagger\left(\epsilon+\frac{\partial_x^2}{2M}\right)\tilde d.
\ee
The crucial point is that the representation of the fermion field now contains a vertex operator:\be
\Psi(x)\sim d^\dagger(x)\sim \tilde d^\dagger(x)e^{-i\sqrt{2\pi \nu}\Theta(x)}, 
\ee
where \be
\Theta(x)=  \tilde\varphi(x)+  \tilde\varphi(-x),\ee
and \be
\nu=\gamma^2/(4\pi ^2K).\label{OCcorrection}\ee

After the unitary transformation, we can calculate   correlations
for the free fields using standard methods.  The Green's function for
the free $\tilde d$ must be calculated with the proper mode expansion
in terms of standing waves, $\tilde d(x)=\sqrt{(2/\pi)}\int_0^{k_0}
dk\, \sin(kx)\tilde d_k$, where $k_0\ll a^{-1}$ is the momentum cutoff
of the impurity sub-band. We obtain
\be  
\langle \tilde d(x,t)\tilde
d^\dagger(x,0)\rangle=e^{-i\epsilon t}\sqrt{\frac{-iM}{2\pi
    (t+i0)}}\left[1-e^{i2Mx^2/(t+i0)}\right].\label{freeimp}
\ee
In the
bulk regime of Eq. (\ref{freeimp}), we neglect the rapidly oscillating
factor $\propto e^{i2Mx^2/t}$; in this case, the free impurity
propagator decays as $\sim t^{-1/2}$. In the boundary case, we expand
for $x\sim a\ll \sqrt{t/M}$ and the free impurity propagator decays as
$\sim t^{-3/2}$. This faster decay is due to the vanishing of the wave
function at the boundary. It can also be understood by noting that at
the boundary the impurity correlator can be calculated as \be \langle
\tilde d(a,t)\tilde d^\dagger(a,0)\rangle\sim a^2 \langle
\partial_x\tilde d(0,t)\partial_x\tilde d^\dagger(0,0)\rangle,\ee and
each spatial derivative amounts to an extra factor of $t^{-1/2}$ due
to the quadratic dispersion of the band-edge mode.

In addition to the free impurity propagator, we have to consider the correlator \cite{1992_Eggert_PRB_46,1995_Fabrizio_PRB_51,1996_Eggert_PRL_76}\be \langle e^{\pm i\sqrt{2\pi \nu } \Theta(x,t)}e^{\mp i\sqrt{2\pi \nu} \Theta(x,0)}\rangle\propto\left|\frac{ x^2}{t^2(4x^2-v^2t^2)}\right|^\nu.\ee
 Thus, in the bulk case ($2x\gg vt$) the correlator  for the the vertex operator adds a factor of $\sim t^{-2\nu}$ to  the decay of the Green's functions. In the boundary case, the factor is $\sim t^{-4\nu}$, a faster decay that stems from the correlation between $\tilde \varphi(x)$ and $\tilde \varphi(-x)$ for $x\sim a$ (whereas these become uncorrelated right- and left-moving bosons in the bulk). Putting the effects together leads to \bea
G_{\text{osc}}(t,a)&\sim& \langle \tilde d(a,\pm t)\tilde d^\dagger(a,0)\rangle \langle e^{ i\sqrt{2\pi \nu } \Theta(a,t)}e^{- i\sqrt{2\pi \nu} \Theta(a,0)}\rangle\nonumber\\
&\sim &e^{\mp i\epsilon t}t^{-\frac32-4\nu},
\eea
(where $\pm$ corresponds to particle/hole impurity) which is the result in Eqs. (\ref{eq:7prime}) and \eqref{eq:10b}.

The scaling dimension of the vertex operator $e^{-i\sqrt{2\pi \nu}\Theta}$ can be related to a phase shift of the low-energy 
modes  due to scattering with the $d$ hole, establishing a connection with the  orthogonality catastrophe  \cite{2006_Pustilnik_PRL_96}. For the integrable model in Eq. (\ref{Hinteracfermion}), the exact phase shift is a simple function of the Luttinger parameter \cite{2008_Pereira_PRL_100}:\be
\gamma=\pi(1-K),\label{exactgamma}
\ee
where the exact Luttinger parameter is for $0\leq V \leq 1$
\be
K=\frac{\pi}{2(\pi-\arccos V)}.\label{exactK}
\ee
The renormalized band edge frequency is\be
\epsilon=\frac{\pi\sqrt{1-V^2}}{2\arccos V}.
\ee 
The exact velocity of the low-energy modes and the effective mass of the impurity are also known: $
v=M^{-1}=\epsilon$ (in units where $a=1$).

In the free fermion limit, a particle tunneling into or out of the
system is restricted to the free or occupied single-particle
states. As is visible in Fig. \ref{fig:regimes} and
Eq.~\eqref{rho0w} the LDOS is then identically zero outside of
the bandwidth set by the dispersion relation. Turning on interactions
allows for tunnelling processes in which the particle leaving or
entering the system excites additional particle-hole pairs. This leads
to a small but nonzero value for the LDOS beyond the threshold energies. The
effect can be included by carefully tracking the regulators in the
Luttinger liquid correlator
\begin{equation}
\langle e^{i \sqrt{2\pi \nu}\varphi(x,t)}e^{-i \sqrt{2\pi \nu}\varphi(x,0)}\rangle \propto [i(vt - i0)]^{-\nu}
\end{equation}
and the impurity correlator in Eq.~\eqref{freeimp}.  At the boundary
and around the band minumum,
the LDOS can for instance be expressed  as
\begin{align}
 & \rho(\omega \approx - \epsilon,a) \sim \int_{-\infty}^{\infty}dt\, \frac{e^{i(\omega+\epsilon)t}}{(vt +i0)^{4\nu}(t-i0)^{\frac{3}{2}}}\nonumber\\
    &\sim[\theta(\omega+\epsilon) - \sin(4\pi \nu) \theta(-\omega-\epsilon)]|\omega+\epsilon|^{\frac{1}{2} + 4 \nu}.
\end{align}
We see that the shoulder ratio of the two-sided singularity is
determined by an interplay of both the impurity and the low-energy propagators. This is
similar, but slightly different than the two-sided singularities within
the continuum of the spectral
function and the dynamic structure factor \cite{2009_Pereira_PRB_79}
for which the shoulder ratio is determined by the exponents for right- and left-movers and the
impurity propagater is just a delta function.

%%%%%%%%%%%%%%%%%%%%%%%%%%%%%%%%%%%%%%%%%%%%%%%%%%%%%%%%%%%%%%%%%%%%%%%%%%%%%%%%%%%%

\section{Other correlation functions \label{sec:other}} 

The   mobile impurity model in Eq. (\ref{eq:Hdensity}) can be used to calculate the exponents in the long-time decay  and  finite-energy singularities of several dynamical correlation functions \cite{2012_Imambekov_RMP_84}. The general recipe  for U(1)-symmetric models is to ({\it i}) identify the operator in the effective field theory that excites the band edge mode and carries the correct quantum numbers; ({\it ii}) write the operator in terms of free impurity and free bosons after the unitary transformation; and ({\it iii}) compute the correlator using the folding trick in the boundary case. In this section we apply this approach to calculate the exponents in the 
density autocorrelation of spinless fermions, spin autocorrelations of spin chains, and the single-particle Green's function of spinful fermions.  

\subsection{Density-density correlation\label{sec:density}}
Let us now consider the density autocorrelation \be
C(t,x)\equiv\langle n(x,t)n(x,0)\label{densityauto} \rangle.  \ee
Using the mode expansion in Eq. (\ref{eq:12}), we obtain the
expression for the density operator including high-energy excitations
\bea
n(x)&=&\Psi^\dagger(x)\Psi(x)\nonumber\\ &\sim &
\psi_R^\dagger\psi_R^{\phantom\dagger}+\psi_L^\dagger\psi_L^{\phantom\dagger}+(e^{i2k_Fx}\psi_L^\dagger\psi_R^{\phantom\dagger}+\text{h.c.})\nonumber\\
&&+\left[(e^{-ik_Fx}\psi_R^\dagger+e^{ik_Fx}\psi^\dagger_L)d^\dagger + \text{h.c.}\right],
\eea
where $k_F=\pi/(2a)$ for the half-filled chain in the model of Eq. (\ref{Hinteracfermion}) and we omitted operators that annihilate the ground state (a
vacuum of $d$ particles). In the boundary case, $\psi_L$ and $\psi_R$
are identified according to Eq. (\ref{foldingpsi}). The leading
operator generated by the low-energy part of $n(x)$ at the boundary is
$\sim \partial_x\varphi(0)$, a dimension-one operator.  As a result, the
LL theory predicts the decay $\langle n(a,t)n(a,0) \rangle\sim
1/t^{2}$. By contrast, in the bulk case the $2k_F$ part of $n(x)$ has
dimension $K$ and gives rise to $\langle n(x\gg a,t)n(x\gg a,0)
\rangle\sim 1/t^{2K}$ as the leading contribution for repulsive
interactions \cite{GiamarchiBOOK}. In summary, the low-energy term
in the density autocorrelation is
\be C_{\text{LL}}(t,x) \sim
t^{-\beta},\label{CLL} \ee with exponents\be \beta_{\text{end}}=2,\qquad
\beta_{\text{bulk}}=2K.  \ee

On the other hand, the high-energy term in the mode expansion for the
density at the boundary yields
\bea
n(a)&\sim
&d^\dagger(a)[e^{-ik_Fa}\psi_R^\dagger(a)-e^{ik_Fa}\psi^\dagger_R(-a)]+
\text{h.c.}\nonumber\\
&\sim &\sin(k_Fa)d^\dagger(a)\psi_R^\dagger(a) + \text{h.c.}.
\eea
After bosonizing and performing the unitary transformation, we find
that the high-energy term  is given by
\bea
n(a)&\sim &  \tilde d^\dagger(a) \exp\left[i\sqrt{\frac{\pi}{2}}\left(\frac{1-\gamma/\pi}{\sqrt{K}}+\sqrt{K}\right)\varphi(a)\right]\times\nonumber\\
&&\times
\exp\left[i\sqrt{\frac{\pi}{2}}\left(\frac{1-\gamma/\pi}{\sqrt{K}}-\sqrt{K}\right)\varphi(-a)\right]
+ \text{h.c.}\nonumber\\
&\sim &a\partial_x\tilde d^\dagger(0)
\exp\left[i\sqrt{2\pi}\left(\frac{1-\gamma/\pi}{\sqrt{K}}\right)\varphi(0)\right]\label{densoperator}\\
&&+ \text{h.c.},\nonumber
\eea
where we kept the leading operator in the expansion of  the slowly-varying fields. From Eq. (\ref{densoperator}) it is straightforward to show that the autocorrelation function  contains a term oscillating with the frequency of the high-energy hole:\be
C_{\text{osc}}(t,x) \sim e^{-i\epsilon t} t^{-\bar\beta },\label{Cosc}
\ee
with the boundary exponent
\be
\bar\beta_{\text{end}}=\frac32+\frac{(1-\gamma/\pi)^2}{K}.
\ee
This should be compared with the corresponding  exponent in the bulk case \cite{2008_Pereira_PRL_100}\be
\bar\beta_{\text{bulk}}=\frac{1+K}{2}+\frac{(1-\gamma/\pi)^2}{2K}.
\ee
Therefore, the exponents associated with the frequency-$\epsilon$
oscillating term in the density autocorrelation obey the relation
\be
2\bar\beta_{\text{bulk}}-\bar\beta_{\text{end}}=K-\frac12.\label{relationeta}
\ee

As mentioned in Section \ref{sec:impuritymodel}, in lattice models   we also have to consider the band-edge mode corresponding to a particle at the top of the band. 
In this case the density operator  contains an additional
 term that creates two high-energy modes, namely a hole at $k=0$ and a particle at $k=\pi$.  In
the noninteracting bulk case of Hamiltonian \eqref{H0}, this term
yields a contribution that behaves as $\sim e^{-i2\epsilon_0t}/t$,
where the slow $1/t$ decay stems from the propagators of the high-energy
particle and hole. However, in the presence of a repulsive interaction
$V>0$ the decay of this contribution changes to $\sim e^{-i2\epsilon
  t}/t^2$ and decays faster than the frequency-$\epsilon$ term for
$t\gg 1/(Ma^2V^2)$  \cite{2008_Pereira_PRL_100}. In the boundary case the equivalent contribution
is subdominant even in the noninteracting case, where it becomes $\sim
e^{-i2\epsilon_0t}/t^3$ due to the faster $t^{-3/2}$ decay of the free
impurity propagator at the boundary. Therefore, the long-time decay of
the density autocorrelation $C(t,x=a)$ is well described by a
combination of the LL term in Eq. (\ref{CLL}) and the
frequency-$\epsilon$ term in Eq. (\ref{Cosc}).

For the integrable model in Eq. (\ref{Hinteracfermion}), we can calculate the exponents $\bar\beta_{\text{bulk}/\text{end}}$ using Eqs. (\ref{exactgamma}) and (\ref{exactK}).   We also note that the power-law decay of $C_{\text{osc}}(t,x)$ implies a finite-energy nonanalyticity  in the Fourier transform  \be
C(\omega,x)\sim |\omega-\epsilon|^{\bar\beta-1}.
\ee

\subsection{Spin autocorrelations\label{sec:spin}}
As an application of our theory to spin chains, we consider the   spin-1/2 XXZ model with an open boundary \be
H_{\text{XXZ}}=\sum_{j\geq 1}\left[\frac{1}{2}(S^+_jS^-_{j+1}+\text{h.c.})+\Delta S^z_jS^z_{j+1}\right],\label{openXXZ}
\ee
where $\mathbf S_j$ is the   spin operator on site $j$ and $\Delta$ is the anisotropy parameter.  We are interested in the long-time decay of the  longitudinal ($\parallel$) and  transverse   ($\perp$)  spin autocorrelations \bea
C^\parallel(t,j)&\equiv&    \langle S_j^z(t)S_j^z(0)\rangle,\\
C^\perp(t,j)&\equiv&    \langle S_j^+(t)S_j^-(0)\rangle.
\eea
We focus on the critical regime $0\leq\Delta\leq 1$. Via a Jordan-Wigner transformation \cite{GiamarchiBOOK} \bea
S^z_j&=&\Psi^\dagger(j)\Psi(j)-\frac12,\label{JWSz}\\
S^-_j&=&(-1)^j\Psi(j)e^{i\pi\sum_{l<j}\Psi^\dagger(l)\Psi(l)},\label{JWSp}
\eea
the XXZ model is equivalent to the spinless fermion model  in Eq. (\ref{Hinteracfermion}) with interaction strength  $V=\Delta$. Thus, for $\Delta=0$ (the XX chain) the model is equivalent to free fermions and some time-dependent correlations can be calculated exactly \cite{1970_Katsura_Physica_46,1995_Stolze_PRB_52}. For $0<\Delta\leq1$  the  LL approach  predicts the asymptotic decay of   nonoscillating terms in the spin autocorrelations \cite{1992_Eggert_PRB_46}:\be
C_{\text{LL}}^\parallel(t,j)\sim t^{-\beta^\parallel}, \qquad C_{\text{LL}}^\perp(t,j)\sim t^{-\beta^\perp}, \label{CLLboth}
\ee
with exponents
\bea
\beta^{\parallel}_{\text{end}}&=&2,\qquad \beta^{\parallel}_{\text{bulk}}=2K,\\
\beta^{\perp}_{\text{end}}&=&\frac1K,\qquad \beta^{\perp}_{\text{bulk}}=\frac{1}{2K}, 
\eea
where the exact Luttinger parameter is given by Eq. (\ref{exactK}) with $V=\Delta$. 
Notice that the exponents for transverse and longitudinal autocorrelations coincide at the SU(2) point $\Delta=1$, where $K=1/2$.

The    high-energy contributions to the spin operator can be obtained
starting from Eqs. (\ref{JWSz}) and (\ref{JWSp}) and employing the
mode expansion for the fermionic field  in Eq. (\ref{eq:12})
\cite{2012_Imambekov_RMP_84}. In the bulk case,  we find
 \bea
S^z_{j=x}&\sim&\tilde d^\dagger(x)\exp\left[i\sqrt{\frac\pi2}\left(\frac{1+K-\gamma/\pi}{\sqrt{K}}\right)\varphi(x)\right]\\
&&\times\exp\left[i\sqrt{\frac\pi2}\left(\frac{1-K-\gamma/\pi}{\sqrt{K}}\right)\varphi(-x)\right]
+ \text{h.c.},\nonumber\\
S_{j=x}^{-}&\sim &\tilde d^\dagger(x)\exp\left[-i\sqrt{\frac\pi2}\left(\frac{K+\gamma/\pi}{\sqrt{K}}\right)\varphi(x)\right]\nonumber\\
&&\times\exp\left[i\sqrt{\frac\pi2}\left(\frac{K-\gamma/\pi}{ \sqrt{K}}\right)\varphi(-x)\right].
\eea
At the boundary, we obtain
  \bea
S_1^{z}&\sim &\partial_x\tilde
d^\dagger(0)\exp\left[-i\sqrt{2\pi}\left(\frac{1-\gamma/\pi}{\sqrt{K}}\right)\varphi(0)\right]\nonumber
\\&&+ \text{h.c.},\\
S_1^{-}&\sim &\partial_x\tilde d^\dagger(0)\exp\left[-i\sqrt{2\pi}\left(\frac{\gamma}{\pi \sqrt{K}}\right)\varphi(0)\right].
\eea
Calculating the correlators along the same lines as the previous examples,  we obtain the oscillating terms in the autocorrelations\bea
C^\parallel_{\text{osc}}(t,j)&\sim &e^{-i\epsilon t}t^{-\bar\beta^{\parallel}},\label{Coscpar}\\
C^\perp_{\text{osc}}(t,j)&\sim &e^{-i\epsilon t}t^{-\bar\beta^{\perp}},\label{Coscperp}
\eea
where \bea
 \bar\beta^{\parallel}_{\text{end}}&=&\frac32+\frac{(1-\gamma/\pi)^2}{K},\\
 \bar \beta^{\perp}_{\text{end}}&=&\frac32+\frac{(\gamma/\pi)^2}{K}.
\eea
We also present, for comparison, the previously known exponents in the bulk \cite{2008_Pereira_PRL_100,2011_Karimi_PRB_84}: \bea
\bar \beta^{\parallel}_{\text{bulk}}&=&\frac{1+K}{2}+\frac{(1-\gamma/\pi)^2}{2K},\\
\bar \beta^{\perp}_{\text{bulk}}&=&\frac{1+K}{2}+\frac{(\gamma/\pi)^2}{2K}.
\eea
The results for the longitudinal spin autocorrelation are the same as those for the density autocorrelation  derived in Section \ref{sec:density}, as expected from the mapping in Eq. (\ref{JWSz}). The bulk and boundary exponents for the  spin autocorrelations obey a relation equivalent  to  Eq. (\ref{relationeta})\be
2\bar\beta^{\perp/\parallel}_{\text{bulk}}-\bar\beta^{\perp/\parallel}_{\text{end}}=K-\frac12,\label{relationetaspin}
\ee
which is independent of $\gamma$.

For the  XXZ model we can simplify the result for the exponents using the exact phase shift in Eq. (\ref{exactgamma}). 
The bulk exponents become 
\bea
 \bar\beta^{\parallel}_{\text{bulk}}&=&K+\frac12,\\
  \bar\beta^{\perp}_{\text{bulk}}&=&K+\frac1{2K}-\frac12.
\eea
Our new results for the boundary exponents are\bea
  \bar\beta^{\parallel}_{\text{end}}&=&K+\frac32,\\
   \bar\beta^{\perp}_{\text{end}}&=&K+\frac1K-\frac12.
\eea

\subsection{Green's function for spinful fermions}

We now consider interacting spin-1/2 fermions, as described by the
Hubbard model \bea
H&=&-\sum_{x\geq1}\sum_{\sigma=\uparrow,\downarrow}[\Psi_\sigma^\dagger(x)\Psi^{\phantom\dagger}_{\sigma}(x+1)+\text{h.c.}]\nonumber\\
&&+U\sum_{x\geq1}n_{\uparrow}(x)n_{\downarrow}(x),\label{hubbard}
\eea
where $U>0$ is the repulsive on-site interaction. Away from
half-filling and in the absence of an external magnetic field, the
low-energy spectrum is described by two bosonic fields corresponding
to decoupled charge and spin collective modes. Our purpose here is to
illustrate the effects of spin-charge separation on finite-energy
contributions to time-dependent correlation functions. We focus on the
single-particle Green's function \be G_\uparrow(t,x)=\langle
\{\Psi^{\phantom\dagger}_\uparrow(x,t), \Psi^{\dagger}_\uparrow(x,0)\}
\rangle .  \ee

In the case of spinful fermions, singular features of dynamic
correlations can in principle   come from both spinon and holon
impurities interacting with the low-energy modes
\cite{2010_Schmidt_PRL_104,2010_Schmidt_PRB_82,2015_Essler_PRB_91}.
For repulsive interactions, the spin velocity is smaller than the charge velocity \cite{GiamarchiBOOK}, so the lower threshold of the
spinon-holon continuum is expected to correspond to a finite-energy
spinon impurity rather than a holon. Here we focus on the contribution from a single high-energy spinon to the Green's function 
and to the LDOS. It is implicitly assumed that the fermion-fermion interactions are strong enough that there is a sizeable separation 
between the spinon band edge and the holon band edge. Otherwise, weak interactions would imply a small energy scale for spin-charge separation, making it difficult to resolve 
the two contributions in real time or in the frequency domain.

%This leads to a threshold in the
%momentum and energy resolved spectral function which, in the bulk, is
%protected by momentum conservation. The spectral weight of the holon
%impurity, on the other hand, appears within a continuum of states and
%the singularity predicted by the impurity model is expected to be
%smoothened by irrelevant operators not taken into account (at least
%when integrability is broken).

%In the case of the Hubbard model, integrability presumably protects
%the holon threshold in the continuum and its effect should be
%noticable in the decay as additional frequecy with powerlaw decay in
%$G_{\uparrow}(t,x)$. For general models with a boundary on the other hand, also the
%spinon impurity may aquire a finite lifetime such that stricly
%speaking there is no powerlaw decay (see Section
%\ref{sec:integrability} for further discussion). The methods discussed here in principle
%allow for the study of all these effects. To be definite, we focus on the
% contribution from the spinon impurity to the  LDOS and Green's
% function, which is expected to be the dominant effect.

We follow the construction in
Ref. \onlinecite{2015_Essler_PRB_91} to define the operators that
create finite-energy spinons coupled to low-energy charge and spin
bosons, maintaining the  correct quantum numbers.
Starting from bosonization
  expressions like
\begin{equation}
\label{eq:1}
\psi_{R,\sigma} \sim e^{-i \sqrt{2\pi}\phi_{R\sigma}},
\end{equation}
we go to a spin and charge separated basis. The physical field is
expanded in right and left movers and written in terms of charge and
spin degrees of freedom. We will only need the right moving component
for which the spinon part is projected onto the impurity
operator. This leads to the projection
\begin{equation}
 \Psi_{\uparrow}\sim d^{\dag}_s e^{-i
   \sqrt{\pi}(\frac{1}{2}\Phi^{*}_{s} -\frac{1}{2}\Phi^{*}_{c} +
     \Theta^{*}_{c})}.
\end{equation}
Here $\Phi^{*}_{\nu}$ and $\Theta^{*}_{\nu}$, with $\nu=c,s$ for charge or spin, respectively,  are the conjugate bosonic
fields that diagonalize the
Hamiltonian at the Luther-Emery point where spin and
charge modes are exactly separated. The bosonic fields satisfy  $[\partial_x \Phi^{*}_{\nu}(x),\Theta_{\nu^\prime}^{*}(x')] = i \delta_{\nu\nu^\prime}\delta(x-x')$.

The impurity model is
\begin{align}
 H_{\text{MIM}} =&  \int_0^{\infty} dx \sum_{\nu = c,s}
 \frac{v_{\nu}}{2}\left[\frac{1}{2K_{\nu}}
   \left(\partial_x\Phi^{*}_{\nu} \right)^2 + 2K_{\nu} \left(\partial_{x}\Theta^{*}_{\nu}\right)^2\right]\nonumber \\
 &+\int_0^{\infty}dx\, d_s^{\dag}\left(\epsilon_s + \frac{\partial_x^2}{2M_s}\right)d_s\nonumber\\
 &+\int_0^{\infty}dx\,\sum_{\nu} \frac{v f_\nu}{\sqrt{\pi }}
 d_s^{\dag}d_s\partial_x\Phi^{*}_\nu,
\end{align}
where $v_{c,s}$ are the charge and spin velocities, respectively, $K_{c,s}$ are the Luttinger parameters, $\epsilon_s$ and $-M_s$ are the energy and effective mass of the high-energy spinon, and  $f_{c,s}$ are impurity-boson coupling constants.  At the Luther-Emery point with free holons and spinons \cite{2015_Essler_PRB_91}, we have $K_c=K_s=1/2$ and $f_c=f_s=0$. In contrast, SU(2)-symmetric models  correspond to  strongly interacting spinons.

We decouple the impurity mode by the unitary transformation
\begin{equation}
 U = \exp\left\{-i \sum_{\nu} \frac{K_{\nu}f_{\nu}}{v_{\nu}\sqrt{\pi}}
 \int_0^{\infty}dx\, d^{\dag}_sd_s\Theta^{*}_{\nu} \right\}.
\end{equation}
We then implement the boundary conditions by the folding trick and
diagonalize the low-energy part of the Hamiltonian by a canonical
transformation. We define
\begin{equation}
 \gamma_{\nu} = \frac{K_{\nu}f_{\nu}}{v_{\nu}}.
\end{equation}
The final expression for the projection of the spinful fermion field operator is
\begin{align}
 \hspace*{-0.5cm}\Psi_{\uparrow}(x)&\sim \tilde{d}^{\dag}_s(x)\exp\left\{ \left(-
 \frac{\sqrt{2K_s}}{4} + \frac{\gamma_s}{\pi
   \sqrt{2K_s}}\right)\varphi_s(x)\right.\nonumber\\
 &+\left( \frac{\sqrt{2K_s}}{4} + \frac{\gamma_s}{\pi
   \sqrt{2K_s}}\right)\varphi_s(-x)\nonumber\\
 &+\left(\frac{1}{2 \sqrt{2K_c}} + \frac{\sqrt{2K_c}}{4} +
 \frac{\gamma_c }{\pi \sqrt{2K_c}}\right) \varphi_c(x)\nonumber\\
 &+\left.\left(\frac{1}{2 \sqrt{2K_c}} - \frac{\sqrt{2K_c}}{4} +
 \frac{\gamma_c }{\pi \sqrt{2K_c}}\right) \varphi_c(-x)\right\}.\label{electron}
\end{align}
Here $\varphi_{c,s}(x)$ represent the free low-energy charge and spin
modes after decoupling of the impurity and $\tilde{d}_s^{\dag}$ creates
the decoupled spinon mode. 

The exponents for the corresponding
oscillating contribution of $G_{\uparrow}(t,x)$ are easily read
off from Eq. (\ref{electron}). Let us restrict ourselves to the SU(2) invariant case
appropriate for the Hubbard model at zero magnetic field. In this case
$K_s=1$ and $\gamma_s = -\pi/2$.  We obtain
\begin{equation}
 G(t,x) \sim e^{-i \epsilon_s t}t^{-\nu^{(s)}},
\end{equation}
with
\begin{align}
 \nu^{(s)}_{\rm bulk} &= 1 + \frac{K_c}{4} + \frac{1}{4K_c}\left(1 +
 \frac{2 \gamma_c}{\pi} \right)^2,\\
 \nu^{(s)}_{\rm end} &= 2  + \frac{1}{2K_c}\left(1 +
 \frac{2 \gamma_c}{\pi} \right)^2. 
\end{align}
The singular behavior of the LDOS is
obtained by Fourier transformation as before. We also obtain the relation
\begin{equation}
  2 \nu^{(s)}_{\rm bulk}-\nu^{(s)}_{\rm end}  =  \frac{K_c}{2},
\end{equation}
which is independent of $\gamma_c$.  It would be interesting to test this
prediction numerically and investigate the relative importance
of the spinon and holon impurity configuration for the autocorrelation and LDOS of the Hubbard model.

%%%%%%%%%%%%%%%%%%%%%%%%%%%%%%%%%%%%%%%%%%%%%%%%%%%%%%%%%%%%%%%%%%%%%%%%%%%%%%%%%%%%

\section{Role of integrability\label{sec:integrability}}

Our results predict the exponents of autocorrelation functions at the boundary of critical one-dimensional systems \emph{assuming} that the long-time decay is described by a power law. By Fourier transform, the same theory predicts the exponent of the nonanalyticity  at the finite energy $\omega=\epsilon$ in the frequency domain. We expect this to hold  for integrable models, where one can calculate a well-defined band-edge frequency from the renormalized dispersion relation (or dressed energy) for the  elementary excitations. Examples of integrable models with open boundary conditions  include the open XXZ chain \cite{1987_Alcaraz_JPA_20,1988_Sklyanin_JPA_21} in Eq. (\ref{openXXZ}) [or, equivalently, its fermionic version in Eq. (\ref{Hinteracfermion})] and the Hubbard model \cite{1985_Schulz_JPhysC_18} in Eq. (\ref{hubbard}), on which many of the previous studies of local spectral properties are based. 

In generic, nonintegrable models, the persistence of a nonanalyticity   inside a multiparticle continuum is questionable. It has been argued that a finite-energy singularity  can be protected in 1D systems by  conservation of quantum numbers in   high-energy bands \cite{2000_Balents_PRB_61}. However, the high-energy subband in our effective mobile impurity model is defined by a projection of the band edge modes, which carry the same quantum numbers as the low-energy modes. Thus, strictly speaking there is no conservation law associated with the number of $d$ particles. 

\begin{figure}
  \begin{center}
\includegraphics*[width=.95\columnwidth]{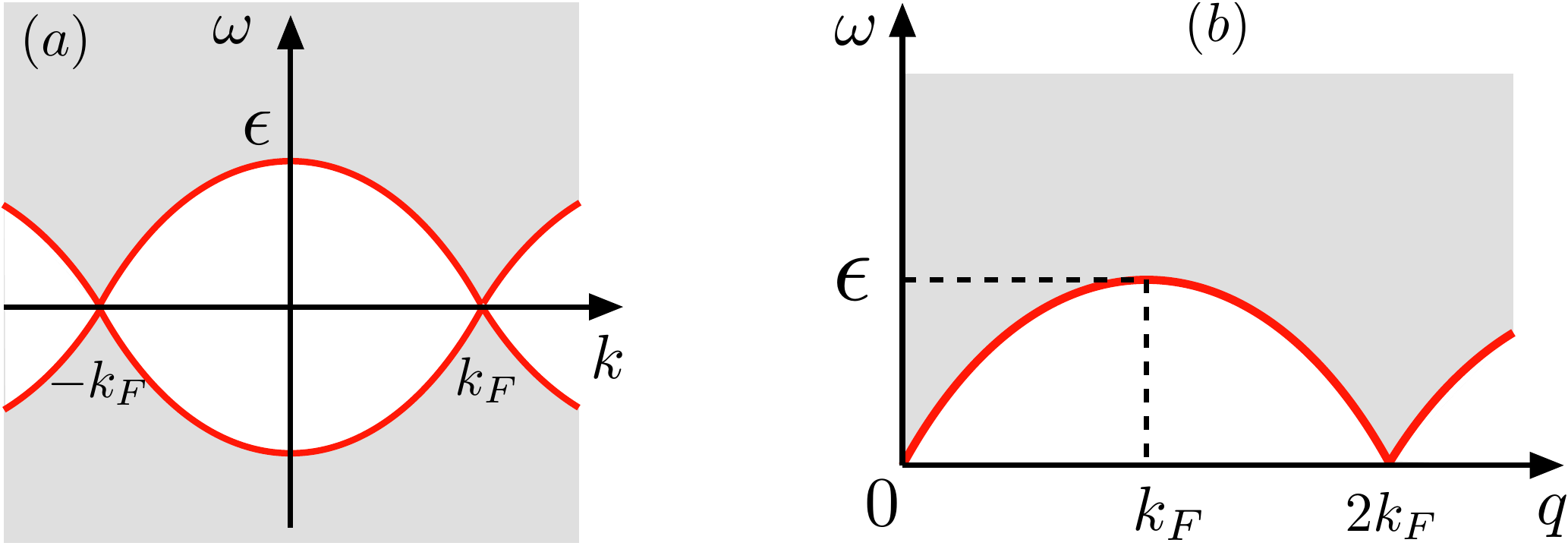}
\end{center}
\caption{(Color online) (a) Support of the single-fermion spectral function $A(k,\omega)$ for a generic 1D model of interacting fermions with Fermi momentum $k_F$. The solid red line represents the lower threshold $\omega_-(k)$, below which $A(k,\omega)$ vanishes. The band edge frequency can be identified as  $\epsilon=\omega_-(k=0)$. (b) Support of the dynamical structure factor $S(q,\omega)$.
}\label{fig:support}
\end{figure}

Nonetheless, we can argue that the band edge  is still well defined for bulk correlations in a semi-infinite system. In the bulk one can measure momentum-resolved response functions, for instance  the spectral function \bea
A(k,\omega)&=&\frac1{2\pi}\int_{-\infty}^{\infty}dt\,e^{i\omega t}\sum_{y}e^{-iky} \nonumber\\
&&\times\langle\{\Psi(x+y,t),\Psi^\dagger(x,0)\}\rangle,\eea
or the dynamical structure factor \be
S(q,\omega)=\frac1{2\pi}\int_{-\infty}^{\infty}dt\,\sum_{y}e^{-iqy}\langle n(x+y,t)n(x)\rangle.
\ee
In momentum-resolved dynamical correlations, the spectral weight vanishes identically below a lower threshold  \cite{2012_Imambekov_RMP_84} [see Fig. \ref{fig:support}(a)]. This threshold is defined by kinematic constraints and exists even for nonintegrable models. The mobile impurity model in the bulk then predicts a power-law singularity  as the frequency approaches the threshold from above. For instance, for the positive-frequency part of the spectral function    \cite{2009_Pereira_PRB_79}:\be
A(k,\omega)\sim [\omega-\omega_-(k)]^{-1+2\nu},
\ee
with $\nu$ defined in Eq. (\ref{OCcorrection}). The band edge frequency that governs the oscillations in local correlations can be identified from the spectrum as a local maximum in the lower threshold, about which the threshold is approximately parabolic. For the spectral function this happens for $k\approx 0$:\be
\omega_-(k\approx 0)\approx \epsilon-\frac{k^2}{2M}.
\ee 
In the dynamical structure factor, the band edge can be read off from the value of the lower threshold  at momentum $q=k_F$, corresponding to the excitation composed of a hole at $k=0$ and a particle at the Fermi point $k=k_F$ [Fig. \ref{fig:support}(b)]. 

The nonanalyticities  in the local bulk correlations are related to  the threshold singularities  of the momentum-resolved correlations by integration over momentum. For instance,  integrating the spectral function implies that  the LDOS behaves as
\bea
\rho(\omega,x\gg a)&=& \int_{-\pi/a}^{\pi/a} dk\, A(k,\omega)\nonumber\\
&\sim &\int_{-k_0}^{k_0} dk\, \theta\left(\omega -\epsilon+\frac{k^2}{2M}\right)\nonumber\\
&&\times\left|\omega -\epsilon+\frac{k^2}{2M}\right|^{-1+2\nu}\nonumber\\
&\sim &|\omega-\epsilon|^{-\frac12+2\nu}.
\eea
Since the singularities in the momentum-resolved dynamic response cannot be broadened, the power-law decay of autocorrelations in the bulk is a generic property of critical 1D systems. 

However, since momentum is not conserved in the presence of a
boundary, the above argument cannot be used to establish power-law
decay of autocorrelation functions at the boundary. From the field
theory perspective, the difference between bulk and boundary cases can
be understood by analyzing the effects of boundary operators that
perturb the mobile impurity model in Eq. (\ref{eq:Hdensity}). In the
following we shall argue that, although formally irrelevant, boundary
operators introduce two important effects in  nonintegrable
models: ({\it i}) they may renormalize the frequency of oscillations
in the boundary autocorrelation, which will then differ from the
frequency in the bulk (only the latter being equal to the band edge
frequency $\epsilon$); ({\it ii}) boundary operators that do not
conserve the number of particles in high-energy subbands may give rise
to a decay rate for the mobile impurity, which implies exponential
decay of the boundary autocorrelation in time and the associated
broadening of the nonanalyticity in the frequency domain.

For discussion purposes we will focus on the regime of weak
interactions, which can be analyzed by perturbation theory in the free
fermion basis, but the argument can be made more general by bosonizing
the low-energy sector and the main points carry through.
If we are interested in the impurity decay, we can furthermore safely neglect
operators that involve the impurity  field but do not couple it to the
low-energy modes---these will at most renormalize the impurity dispersion.

As a simple
example of a boundary operator respecting the symmetries and boundary
conditions, consider the
impurity-number-conserving perturbation
\begin{equation}
 \partial H = g \partial_{x}d^{\dag}(0)\partial_xd(0)\psi^{\dag}(0)\psi(0). 
\end{equation}
Here we use $\psi(x) = \psi_R(x) = -\psi_L(-x)$ to denote the
low-energy modes of the fermion field on the unfolded line.
We will assume that $\partial H$ is present in the effective Hamiltonian and
analyze its influence on the impurity propagator in perturbation
theory.

It is convenient to Fourier transform the time coordinate to make use
of energy conservation, but not the space coordinate. We can organize the
diagrammatic expansion of the time-ordered impurity propagator
\begin{equation}
  G_d(x,x';t) = \langle T d(x,t)d^{\dag}(x',0)\rangle 
\end{equation}
using the Dyson equation
\begin{multline}
  G_d(x,x;\omega) = G_d^{(0)}(x,x;\omega)\\
  + \int dx_1 \int dx_2 G_d^{(0)}(x,x_2;\omega)\Sigma(x_2,x_1;\omega)G_d(x_1,x;\omega).
\end{multline}
If we take only boundary operators into account,
the self-energy $\Sigma$ is purely local:
\begin{equation}
 \Sigma(x_2,x_1;\omega)  = \Sigma(\omega) \delta(x_1 - a)\delta(x_2 -a).
\end{equation}
The solution of the Dyson equation for $x_1=x_2=a$ is
\begin{equation}
 G_d(a,a;\omega)= \frac{1}{[G_d^{(0)}(a,a;\omega)]^{-1} - \Sigma(\omega)  }.\label{dysonresult}
\end{equation}
It follows from  Eq. (\ref{dysonresult}) that the non-analyticity in the LDOS  will be broadened if the local self-energy $\Sigma(a,a;\omega)$ has a nonzero  imaginary part at $\omega=\epsilon$.

For the continuation of this calculation, let us use the notation $G(t) = G(a,a;t)$ for boundary
propagators. The free propagator for the $d$-particle at the boundary
is
\begin{equation}
 G^{(0)}_d(t) = \frac{(-iM)^{3/2}}{\sqrt{2\pi}}\frac{\theta(t)e^{-i\epsilon t}}{(t+i \eta/v)^{3/2}},
\end{equation}
while for the low energy modes we have
\begin{equation}
  G^{(0)}_{LL}(t) \equiv \langle T \psi(a,t)\psi^{\dag}(a,0)\rangle = [2\pi i( vt - i \eta \sgn t)]^{-1},
\end{equation}
 where $\eta$ is a short-distance cutoff and is related to the bandwidth of the impurity and low-energy subbands.

 The first order correction in the coupling constant $g$ corresponds
to a tadpole diagram proportional to the density of low-energy modes at
 the boundary.  It will not induce the decay rate that we are after
 [rather, it is like a nonuniversal renormalization of the coupling constant of the boundary operator $\partial_xd^\dagger(0)\partial_xd(0)$, which does not couple the impurity to the low-energy modes].
The second order correction
 is given by the expression
\begin{align}
  &\delta \Sigma^{(2)}(\omega) =-i g^2\int_{-\infty}^{\infty} dt\, e^{i\omega t}G^{(0)}_{LL}(t)G^{(0)}_{LL}(-t)G^{(0)}_d(t).
  %\\  &=   \left(\frac{ g}{2\pi}\right)^2 \frac{(-im)^{3/2}}{\sqrt{2\pi}} \int_0^{\infty}dt \frac{i \theta(t) e^{i(\omega-\epsilon)t}}{[vt - i\eta][vt + i\eta]t[t+i\eta]^{1/2}}\nonumber
\end{align}
%% [see Fig.~\ref{fig:self-energy}]
The imaginary part is then obtained as
\begin{multline}
  \im \delta \Sigma^{(2)} = -\left(\frac{ g}{2\pi}\right)^2 \frac{M^{3/2}}{\sqrt{\pi}}\\
  \times\int_{-\infty}^{\infty}dt  \frac{ e^{i(\omega-\epsilon)t}}{(vt - i\eta)(vt + i\eta)(t+i\eta/v)^{3/2}}.\label{ImSigmairrelevant}
\end{multline}
By power counting in the integral we see that
\begin{equation}
 \delta \Sigma^{(2)}(\omega) \propto |\epsilon - \omega|^{5/2},
\end{equation}
and hence the self energy vanishes on-shell, when $\omega =
\epsilon$, so this correction will not induce a finite decay rate
\begin{equation}
 \frac{1}{\tau} = - \im \Sigma(\omega = \epsilon).
\end{equation}
The factor of $e^{i(\omega-\epsilon)t}$  in Eq. (\ref{ImSigmairrelevant})   is general for self-energy contributions  generated by 
perturbations that conserve the number of $d$-particles. Therefore, the decay rate must vanish to all orders if, for some reason, the irrelevant interactions conserve the number of high-energy excitations \cite{2000_Balents_PRB_61}. 

To derive a nonzero decay rate, we will have to consider
perturbations that do not preserve the number of impurity modes and may contribute to the self-energy for $\omega=\epsilon$. As
stated before, this is a typical effect of the boundary  breaking
translational invariance, since in the bulk kinematic constraints associated with momentum and energy conservation 
prevent the decay of the band-edge mode. Due to  the U(1) symmetry (conservation of the total charge), the annihilation  (creation) of  a high-energy hole 
entails the  annihilation (creation) of a particle in a low-energy  state.  A family of such boundary operators that are   allowed by symmetry and the boundary
conditions are for example
\begin{equation}
 \partial H_n = g_n \partial_xd(0)[\psi^{\dag}(0)\psi(0)]^n\psi(0) + \text{h.c.}.
\end{equation}
The first nontrivial correction to the self-energy is of second order 
in the coupling $g_n$.% [Fig.~\ref{fig:self-energy}].
The diagram corresponds  to a simple
low-energy propagator dressed by $n$ particle-hole pairs,
\begin{equation}
  \delta \Sigma_n^{(2)}(\omega) = -ig_n^2 \int_{-\infty}^{\infty}dt\, e^{i\omega t} [G^{(0)}_{LL}(t)]^{n +1}[G^{(0)}_{LL}(-t)]^{n},
\end{equation}
leading to\be 
  \im \delta \Sigma_n^{(2)}(\omega) = - \frac{g_n^2}{(2\pi v)^{2n+1}}
  \int_{-\infty}^{\infty}dt \frac{ te^{i\omega t}}{i(t^2+\eta^2/v^2)^{n+1}}.
\ee
Closing the contour in the upper half plane and picking up the pole at $t = i \eta/v$, we obtain a cutoff-dependent decay rate
\begin{equation}
 \frac{1}{\tau} \propto g_n^2 e^{-\epsilon \eta/v}.
\end{equation}
In contrast to the earlier case, we do find a possibly finite decay 
rate. We note that $ \epsilon\eta/v\sim \mc O(1)$  if the short-distance   is of the order of the lattice spacing $a$, but $\epsilon\eta/v\gg 1$ if $\eta\gg a$.   

Boundary operators like   $\partial H_n$ will in principle be
generated from lower order processes for a generic model when we
integrate out the states outside of our impurity and low-energy
subbands in a renormalization group procedure. Physically, we can
think of these processes as the result of a cascade, or particle shower \cite{1934_Bethe_PRSL_856,1938_Landau_PRSL_166},  involving
many intermediate states which are no longer in the description.  The
number $n$ of low-energy particle-hole pairs roughly reflects the
number of microscopic interaction processes and has to be sizeable (of
the order of \(\sim v\eta^{-1}\epsilon^{-1}\)) to accommodate for the
excess energy.  The coupling $g_n$, therefore, will scale with high powers of
the microscopic interaction strength and thus will be very small for
weak interactions leading to a negligible decay rate. Stronger interactions,
however, may show sizeable renormalization effects in the decay rate and
frequency shift of correlations at the boundary.

Coming back to integrability, we argue that the above corrections do
not occur for models with open boundary conditions solvable by Bethe
ansatz.  The argument relies on the fact that the exact eigenstates of
the model still define a conserved impurity state corresponding to a
hole in the quantum number configuration of the ground state. This
state is parametrized by a rapidity $\lambda$ and has well-defined
energy given by the dressed energy function $\epsilon(\lambda)$. One can
in fact show, using the thermodynamic Bethe ansatz, that the spectrum is
still determined by the bulk dressed energy function by a similar type
of folding trick to the one we used for the low-energy theory
\cite{ZvyaginBOOK}. Not only does this imply the absence of a decay rate,
also the impurity energy does not renormalize and the same frequency
should be observed in the autocorrelation in the bulk and at the boundary. The ``miracle'' of integrability thus manifests itself as a
fine tuning of the coupling constants in the effective field theory, 
in this case the vanishing of the couplings $g_n$.

%%%%%%%%%%%%%%%%%%%%%%%%%%%%%%%%%%%%%%%%%%%%%%%%%%%%%%%%%%%%%%%%%%%%%%%%%%%%%%%%%%%%

\section{Numerical results for spin chains\label{sec:tDMRG}} 

In this section, the field theoretical  prediction for the asymptotic behavior of the autocorrelations $C^{\parallel/\perp}(t,j)$
are checked, numerically, for critical spin chains with size $L=300$
and  open boundary conditions. We use the adaptive tDMRG \citep{2004_White_PRL_93,2004_Daley_JSTAT}
keeping up to $m=300$ ($m=450$) states per block for the chains with spin
$S=1/2$ ($S=1$ and $S=3/2$).
The time evolution was performed with the second order Suzuki-Trotter decomposition with time
step $0.025\leq\delta t\leq0.3$. The discarded weight was typically
about $10^{-8}$--$10^{-12}$ during the time evolution.
The numerical error sources in the tDMRG have two origins:
\begin{enumerate}
\item\label{item:1}  The Trotter error, which is related with the order ($n$)
 of the Suzuki-Trotter  decomposition. For the order $n$, this error 
 is of the order $(\delta t)^{n+1}$.
\item\label{item:2} The truncation error associated with the number of discard states. 
\end{enumerate}
These errors can be controlled by decreasing the time step ($\delta t$) and
increasing the number of states kept in the DMRG simulation.

We are interested in the   long-time behavior of the longitudinal and transverse spin autocorrelations at the end site, 
$C_{\text{end}}^{\parallel/\perp}(t) =C^{\parallel/\perp}(t,1)$, 
and  in the bulk, $C_{\text{bulk}}^{\parallel/\perp}(t) = C^{\parallel/\perp}(t,L/2)$.
As discussed in the Section \ref{sec:spin},
these autocorrelations can be described by a combination of universal power laws  predicted by the LL theory and oscillating terms  predicted by the nLL theory. 

\subsection{Integrable spin-$1/2$ model}
First, we consider the integrable spin-1/2 XXZ model in Eq. (\ref{openXXZ}). According to  Eqs. (\ref{CLLboth}), (\ref{Coscpar}), and (\ref{Coscperp}), the real parts of the autocorrelations behave as
\bea
\hspace*{-0.8cm}\text{Re}\left[C_{\text{end}}^{\parallel}(t)\right]&=&\frac{A^{\parallel}_{1}}{t^{2}}+\frac{A^{\parallel}_{2}\cos(Wt+\varphi)}{t^{\frac{3}{2}+\xi}},\label{eq:clongend}\\
\hspace*{-0.8cm}\text{Re}\left[C_{\text{bulk}}^{\parallel}(t)\right]&=&\frac{B^{\parallel}_{1}}{t^{2}}+\frac{B^{\parallel}_{2}}{t^{2\xi}}+\frac{B^{\parallel}_{3}\cos(Wt+\varphi)}{t^{\frac{1}{2}+\xi}}\nonumber\\&&+\frac{B^{\parallel}_{4}\cos(2Wt+\tilde{\varphi})}{t^{\zeta}},\label{eq:clongbulk}\\
\hspace*{-0.8cm}\text{Re}\left[C_{\text{end}}^{\perp}(t)\right]&=&\frac{A^{\perp}_{1}}{t^{\frac{1}{\xi}}}+\frac{A^{\perp}_{2}\cos(Wt+\varphi)}{t^{\xi+\frac{1}{\xi}-\frac{1}{2}}},\label{eq:ctransend}\\
\hspace*{-0.8cm}\text{Re}\left[C_{\text{bulk}}^{\perp}(t)\right]&=&\frac{B^{\perp}_{1}}{t^{\frac{1}{2\xi}}}+\frac{B^{\perp}_{2}}{t^{2}}+\frac{B^{\perp}_{3}\cos(Wt+\varphi)}{t^{\xi+\frac{1}{2\xi}-\frac{1}{2}}}.\label{eq:ctransbulk}
\eea
Here we have imposed the constraint that for the XXZ model the interaction
dependence of all exponents (bulk or boundary, low-energy or high-energy) can
be expressed in terms of a single parameter $\xi$.  The theoretical
prediction is $\xi=K=\frac{\pi}{2\left(\pi-\arccos\Delta\right)}$. The
frequency of the oscillating terms is predicted to be the same for bulk and
boundary autocorrelations, and is given by
$W=\epsilon=\frac{\pi\sqrt{1-\Delta^{2}}}{2\arccos\Delta}$. In
Eq. (\ref{eq:clongbulk}) we included the 
oscillating term with frequency $2W$ which comes from a hole
at $k=0$ and a particle at $k=\pi$ \citep{2008_Pereira_PRL_100}. The corresponding  exponent is predicted to be $\zeta=1$ for $\Delta=0$   but $\zeta=2$ for $0<\Delta<1$ and $t\gg 1/\Delta^2$. In the following we shall test  the analytical predictions from the nLL theory by fitting  the tDMRG data to  the expressions above. 

\begin{figure}
\begin{centering}
\includegraphics[width=7.5cm]{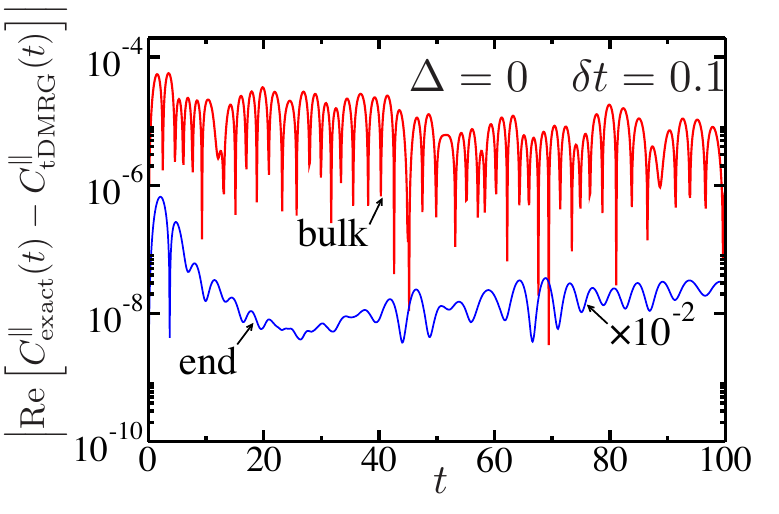}
\par\end{centering}

\caption{(Color online)
The differences between the real parts of the exact results {[}Eq. (\ref{eq:szszexac}){]}
and the tDMRG data for the autocorrelations $C^{\parallel}(t,j)$ 
for the spin-1/2 XXZ chain with $L=300$ and $\Delta=0$. The bulk
(end) case corresponds to $j=L/2$ $(j=1)$. We use $m=200$ DMRG
states and time step $\delta t=0.1$. We multiply the results of $C_{\text{end}}^{\parallel}(t)$
by $10^{-2}$ in order to see both data in the same figure. \label{fig:diff}}
\end{figure}

Before   presenting the fit results, let us   consider the 
 chain  with  $\Delta=0$. At this point, the autocorrelation $C^{\parallel}(t,j)$ 
is equivalent to the  density autocorrelation for free spinless fermion (see Section \ref{sec:GF}). It is straightforward  to show that for even size $L$
\begin{equation}
C^{\parallel}(t,j)=\left[\frac{2}{L+1}\sum_{m=1}^{L/2}\sin^{2}\left(\frac{m\pi j}{L+1}\right)e^{i\varepsilon_mt}\right]^{2},\label{eq:szszexac}
\end{equation}
where $\varepsilon_m=-\cos\left(\frac{\pi m}{L+1}\right)$. In Fig. \ref{fig:diff}, we present the differences between the exact
results of $C_{\text{end}/\text{bulk}}^{\parallel}(t)$ and the tDMRG data obtained
considering $m=200$ and $\delta t=0.1$. As we can see, the agreement
is quite good. It is interesting to note that the errors are of the
order $\sim10^{-4}-10^{-6}$, which are smaller than the errors due
to the use of the second order Suzuki-Trotter decomposition, of order  $(\delta t)^{3}=10^{-3}$. 

The results depicted in Fig. \ref{fig:diff}
 show that we obtain accurate results for the $C_{\text{end}/\text{bulk}}^{\parallel}(t)$
with the tDMRG by using $m=200$ and $\delta t=0.1$. Away from the
point ${\Delta=0}$, we do not have exact results to compare with. In this
case, we compare the autocorrelations $C_{\text{end}/\text{bulk}}^{\parallel/\perp}(t)$
for different values of $m$ ($m=100$, $m=200$ and $m=300$) and
time step $\delta t$ ($\delta t=0.3,$ $\delta t=0.1$, and $\delta
t=0.025)$, in order to estimate the numerical errors. Overall, we
estimate that these errors are at least one order of magnitude smaller
than the values of the autocorrelations acquired by tDMRG.

\begin{figure}

\begin{centering}
\includegraphics[width=4.3cm]{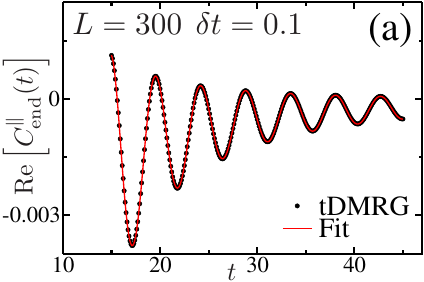}\,\includegraphics[width=4.3cm]{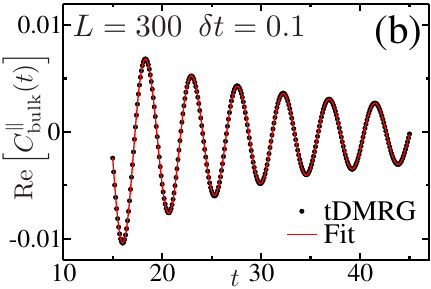}
\par\end{centering}

\begin{centering}
\includegraphics[width=4.3cm]{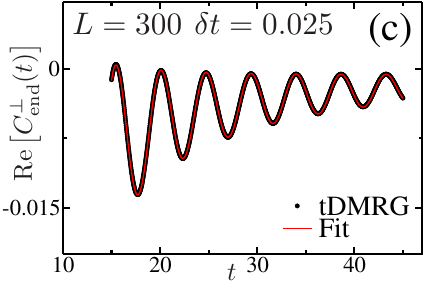}\,\includegraphics[width=4.3cm]{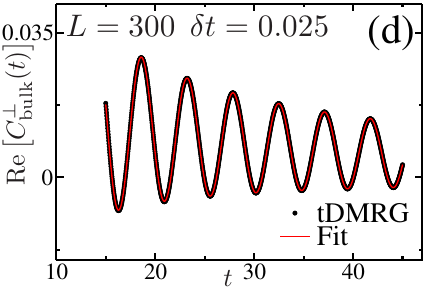}
\par\end{centering}
\caption{(Color online)
Real parts of the autocorrelations $C_{\text{end}/\text{bulk}}^{\parallel/\perp}(t)$ vs. $t$
for the spin-$1/2$ XXZ  chain for $\Delta=0.6$, $L=300$, and
$m=200$. For the longitudinal {[}figures (a) and (b){]} and transverse
{[}figures (c) and (d){]} spin autocorrelations we use $\delta t=0.1$
and $\delta t=0.025$, respectively. The symbols are the tDMRG results
and the solid lines are fits to our data using Eqs. (\ref{eq:clongend})-(\ref{eq:ctransbulk})
(see text). \label{fig:spincorrelations}}
\end{figure}

\begin{figure}

\begin{centering}
\includegraphics[width=6.8cm]{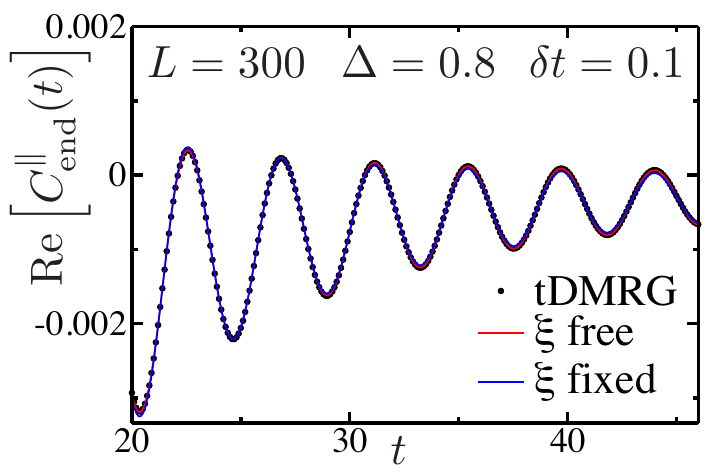}
\par\end{centering}

\caption{(Color online)
Real part of the longitudinal spin autocorrelation $C_{\text{end}}^{\parallel}(t)$ vs. $t$ for the
spin-$1/2$ XXZ chain with anisotropy $\Delta=0.8$ and system size
$L=300$. The data were obtained using $m=200$ DMRG states and time
step $\delta t=0.1$. We fit the tDMRG data to Eq. (\ref{eq:clongend})
taking the parameter $\xi$  to be either free or fixed as $\xi=K$ 
(see legend).\label{fig:fitfixed}}
\end{figure}

Some typical examples of the numerical data fitted to   Eqs. (\ref{eq:clongend})-(\ref{eq:ctransbulk})
are presented in Fig. \ref{fig:spincorrelations} for the spin-1/2
XXZ chain with anisotropy $\Delta=0.6$. The parameters $\xi$
and $W$ obtained by this fitting procedure  are given in Table \ref{tab:parameters}
for some values of the anisotropy $\Delta$. Overall,  the parameters obtained are in agreement
with the theoretical prediction  presented in the last column
of   Table \ref{tab:parameters}. In the fitting procedure, the tDMRG
data considered were in the range $15<t<80$. We note that the parameter
$\xi$ changes slightly depending on the time range used in the fit. One of the
largest discrepancies found corresponds to the parameter $\xi$ obtained
from $C_{\text{end}}^{\parallel}(t)$ for $\Delta=0.8$ (see Table \ref{tab:parameters}).
Although this exponent ($\xi=0.459$) differs slightly from the
predicted ($K=0.6287$), we found a very good agreement of the fit
of the tDMRG data to Eq. (\ref{eq:clongend}) if we consider $\xi=K$
fixed, as shown in Fig. (\ref{fig:fitfixed}). It is also interesting
to note that, even though for some values of $\Delta$ the fitted value of 
$\xi$ is not so close to the predicted one, we found that $|2\beta_{\text{bulk}}^{\parallel/\perp}-\beta_{\text{end}}^{\parallel/\perp}-K+1/2|<0.06$,
which is close to zero in agreement with the relation predicted in  Eq. (\ref{relationetaspin}).

\begin{table}
\caption{(Color online)
The exponent $\xi$ and the band edge frequency $W$ for the autocorrelations
$C_{\text{end}/\text{bulk}}^{\parallel/\perp}(t)$ for the spin-1/2 XXZ chain for
some values of $\Delta$. The parameters $\xi$ and $W$ were obtained
by fitting the tDMRG data to Eqs. (\ref{eq:clongend})-(\ref{eq:ctransbulk}).
The last column are the theoretical predictions for these parameters.\label{tab:parameters}}

\centering{}%
\begin{tabular}{lcccccc}
\hline 
 &  & $C_{\text{end}}^{\parallel}$ & $C_{\text{bulk}}^{\parallel}$ & $C_{\text{end}}^{\perp}$ & $C_{\text{bulk}}^{\perp}$ & Exact\tabularnewline
\hline 
\hline 
\multirow{2}{*}{$\Delta=0$$\quad$} & $\xi$$\quad$ & 0.992$\:$ & 1.006$\:$ & 0.943$\:$ & 0.981$\:$ & 1$\:$\tabularnewline
 & $W$$\quad$ & 1.002$\:$ & 1.000$\:$ & 1.000$\:$ & 1.002$\:$ & 1$\:$\tabularnewline
\hline 
\multirow{2}{*}{$\Delta=0.3$$\quad$} & $\xi$$\quad$ & 0.849$\:$ & 0.829$\:$ & 0.836$\:$ & 0.893$\:$ & 0.8375$\:$\tabularnewline
 & $W$$\quad$ & 1.182$\:$ & 1.183$\:$ & 1.184$\:$ & 1.186$\:$ & 1.1835$\:$\tabularnewline
\hline 
\multirow{2}{*}{$\Delta=0.6$$\quad$} & $\xi$$\quad$ & 0.677$\:$ & 0.678$\:$ & 0.711$\:$ & 0.595$\:$ & 0.7093$\:$\tabularnewline
 & $W$$\quad$ & 1.355$\:$ & 1.355$\:$ & 1.356$\:$ & 1.358$\:$ & 1.3551$\:$\tabularnewline
\hline 
\multirow{2}{*}{$\Delta=0.8$$\quad$} & $\xi$$\quad$ & 0.459$\:$ & 0.554$\:$ & 0.649$\:$ & 0.585$\:$ & 0.6287$\:$\tabularnewline
 & $W$$\quad$ & 1.466$\:$ & 1.465$\:$ & 1.467$\:$ & 1.468$\:$ & 1.4646$\:$\tabularnewline
\hline 
\end{tabular}
\end{table}

\subsection{Effects of bound states and nearly flat bands\label{sec:boundstates}} 

Before we start   analyzing nonintegrable models, let us briefly
describe some situations where the   predictions of Section \ref{sec:spin} do not hold. As  mentioned in Section \ref{sec:impuritymodel}, our  mobile impurity model     assumes that a single type of high-energy excitation (the deep hole) is sufficient to describe the oscillations in the autocorrelation functions. This is equivalent to assuming that   in the frequency domain the dominant finite-energy nonanalyticity occurs  at the   band edge of single-hole excitations. However, more generally   dynamical correlation functions may contain additional singularities at frequencies corresponding to bound states which are absent in the noninteracting model. In this case, additional oscillating components in the long-time decay of $C_{\text{end}/\text{bulk}}^{\parallel/\perp}(t)$ can arise and decay more slowly than the contribution  considered in Eqs. (\ref{eq:clongend})-(\ref{eq:ctransbulk}).  While bound states can be incorporated   in a more general mobile impurity model \cite{2009_Pereira_PRB_79},  in this work we look for examples where the existence of bound states can be ruled out, so we can test the bulk versus boundary behavior of the band edge contribution.  

\begin{figure}
\includegraphics[width=8cm]{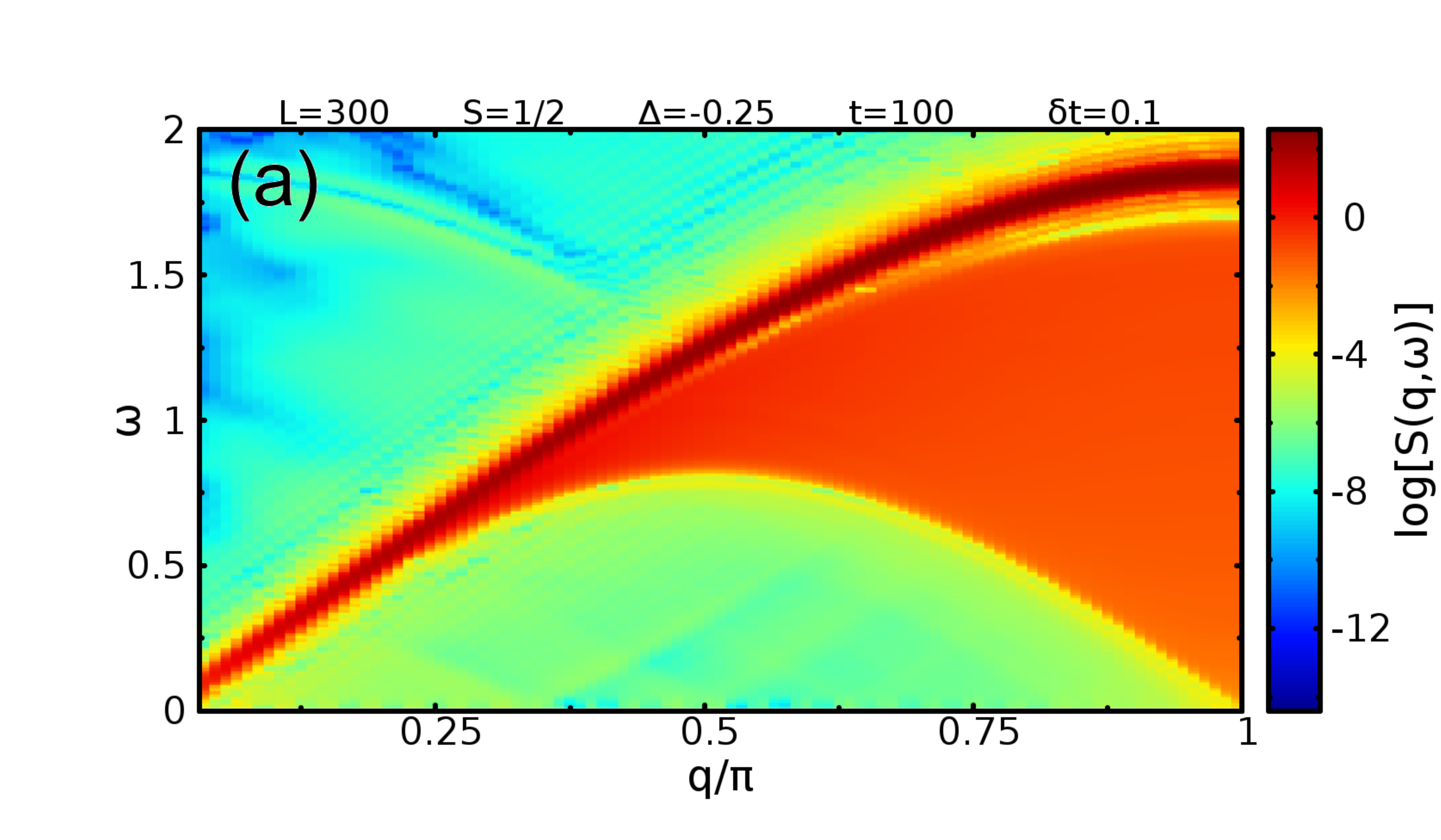}

\includegraphics[width=6.5cm]{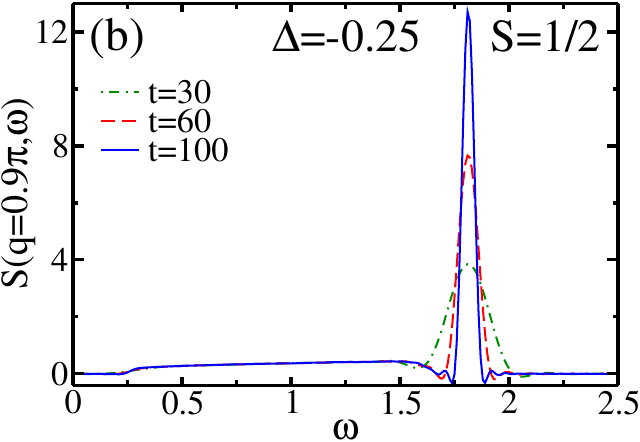}

\caption{(Color online)
(a) Longitudinal spin structure factor for the spin-1/2 XXZ chain with  anisotropy $\Delta=-0.25$
and system size $L=300$. The data were obtained using $m=200$ and
$\delta t=0.1$.  (b) Lines shapes of $S(q=0.9\pi,\omega)$ obtained for different maximum times.\label{fig:boundstate}}
\end{figure}

The signature of bound states can be observed in the longitudinal spin structure factor\be
 S^{\parallel}(q,\omega)=\frac{1}{2\pi}\int_{-\infty}^{\infty} dt\, e^{i\omega t}\sum_je^{-iqj}C^\parallel(t,j).\ee
It is known \cite{2008_Pereira_PRL_100} that for the spin-1/2 XXZ  chain with $-1<\Delta<0$, which is in the critical regime but is equivalent to spinless fermions with \emph{attractive} interactions,  $S^{\parallel}(q,\omega)$ exhibits  a narrow peak above the two-spinon continuum.  This peak  can be interpreted within the effective field theory as a bound state of a high-energy particle and a high-energy hole. Fig. \ref{fig:boundstate} shows $S^{\parallel}(q,\omega)$ for $\Delta=-0.25$. Although this bound state  is inside a continuum of multiple particle-hole pairs, we expect that for the integrable model  the peak in the longitudinal spin structure factor is not broadened by decay processes and is  given by
a delta function, \emph{i.e.}, $S^\parallel(q,\omega)\sim\delta\left(\omega- \Omega_{\text{bs}}(q)\right)$, where $\Omega_{\text{bs}}(q)$ is the dispersion relation of the bound state.
In our numerical results we observe that the peak has a finite width because the frequency resolution is limited by the 
finite time in the tDMRG data. However, as shown in Fig. \ref{fig:boundstate}(b), $S^\parallel(q,\omega)$
becomes narrower as the time increases. This is  a strong evidence of
the existence of a bound state in the spectrum.

Another situation that limits the applicability of our mobile impurity model is when the excitation spectrum contains particles with a large effective mass $M$, {\it i.e.} in the presence of nearly flat bands. As discussed in Section \ref{sec:impuritymodel}, the  exponents of the oscillating terms hold for large times compared to the inverse of the band curvature energy scale,  in the regime $t\gg Ma^{2}$. If the mass   is large, the asymptotic behavior will only be observed  after extremely long times, beyond the reach of the tDMRG method. 

\subsection{Higher-$S$ spin chains}

With the above limitations in mind, we turn to the study of autocorrelations
in  nonintegrable models. In principle, a simple way to break the
integrability of the spin-1/2 XXZ chain (while preserving a gapless spectrum
as well as  U(1) and discrete symmetries) is to add small
next-nearest-neighbor exchange couplings, {\it e.g.}, $\delta H\sim
\sum_jS^z_jS^z_{j+2}$. 
However, it is well known that the adaptive tDMRG only works efficiently for models with
nearest-neighbor exchange couplings \cite{2005_Feiguin_PRL_72}.
For this reason,  we study critical spin-$S$ chains with $S>1/2$ \citep{1983_Haldane_PLA_93,1983_Haldane_PRL_50,1986_Schulz_PRB_34,1992_Alcaraz_PRB_46} as examples of nonintegrable models. We consider the Hamiltonian  \be
H=\sum_{j=1}^L[S^x_jS^x_{j+1}+S^y_jS^y_{j+1}+\Delta S^z_jS^z_{j+1}+D(S_j^z)^2],\label{higherS}
\ee
where $\mathbf S_j$ is the  spin-$S$ operators acting on site $j$, $\Delta$ is the exchange anisotropy and $D$ is the single-ion anisotropy. 

The expressions for  spin-$S$ operators within  the low-energy effective field theory can be obtained by noting that spin chains with $S=n/2$ can be represented by $n$-leg ladders in the limit where strong rung couplings select the spin-$S$ multiplet of the local spins 1/2 \cite{1985_Timonen_JPC_18,1986_Schulz_PRB_34}. For instance, for $S=1$ we   can write $\mathbf S_j=\boldsymbol\sigma_j+\boldsymbol\tau_j$, where $\boldsymbol\sigma_j$ and $\boldsymbol \tau_j$ are two spin-1/2 operators that commute with each other, and use the Jordan-Wigner transformation [essentially two copies of  Eqs. (\ref{JWSz}) and (\ref{JWSp})] to write $\boldsymbol\sigma_j$ and $\boldsymbol \tau_j$ in terms of two fermions, say $\Psi_{\sigma}(j)$ and $\Psi_{\tau}(j)$.  %$\boldsymbol\sigma_j$ and $\boldsymbol \tau_j$  in terms of two fermionic species:\bea
%\sigma^z_j&=&a^\dagger_ja^{\phantom\dagger}_j-1/2,\\
%\sigma^-_j&=&(-1)^ja^{\phantom\dagger}_je^{i\pi \sum_{l<j}a_l^\dagger a^{\phantom\dagger}_l},\\
%\tau^z_j&=&c^\dagger_jc^{\phantom\dagger}_j-1/2,\\
%\tau^-_j&=&(-1)^jc^{\phantom\dagger}_je^{i\pi \sum_{l<j}c_l^\dagger c^{\phantom\dagger}_l}.
%\eea
The resulting fermionic model  turns out to be  strongly interacting (and contain long-range interactions), but the low-energy sector can be treated by bosonization and a renormalization group analysis \cite{1985_Timonen_JPC_18,1986_Schulz_PRB_34}. A critical phase with central charge $c=1$ (analogous to the spin-1/2 XXZ model with $|\Delta|<1$) can be  understood as the result of gapping out all branches of excitations except for one remaining gapless mode. 

Here, we go beyond the low-energy regime and apply the  nLL theory to investigate spin autocorrelations in  the critical phase of model (\ref{higherS}).  Our main goal is to test the predictions of Section \ref{sec:integrability}, namely the frequency shift and exponential decay of oscillating terms in the boundary autocorrelation for nonintegrable models.  In the bulk case, the mobile impurity model of the nLL theory can be applied phenomenologically \cite{2012_Imambekov_RMP_84} after identifying the thresholds of the spectrum in dynamical spin structure factors. Unlike the spin-1/2 XXZ model, however, the coupling between the impurity and the  low-energy modes is not known exactly and is regarded as a phenomenological parameter. 

\begin{figure}

\includegraphics[width=8cm]{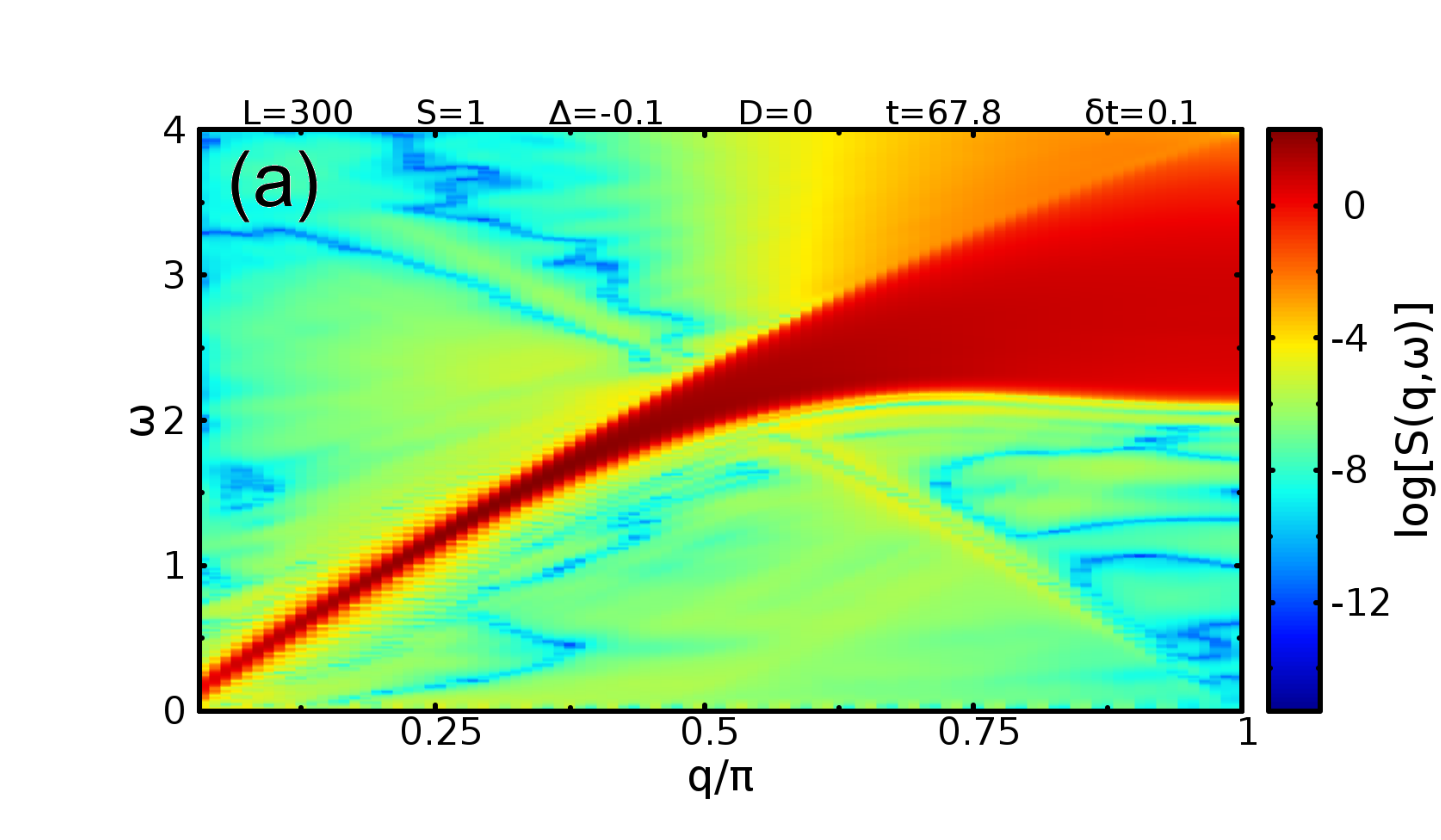}

\includegraphics[width=8cm]{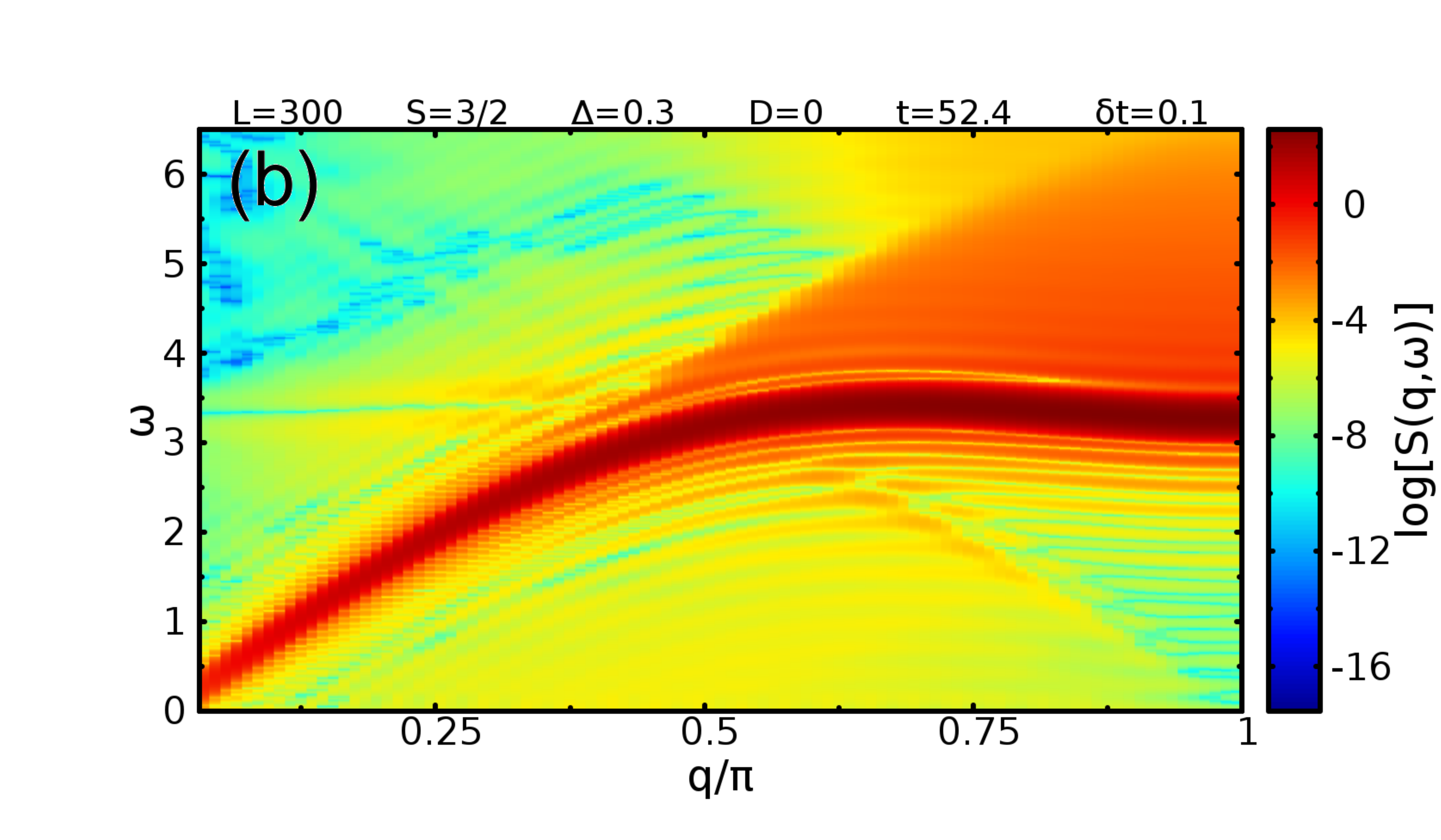}

\caption{(Color online)
The longitudinal spin structure factor of the critical spin-$S$
XXZ chains. (a) Results for $S=1$ and $\Delta=-0.1$ and (b) for
$S=3/2$ and $\Delta=0.3$.\label{fig:Sspins}}
\end{figure}

As our first attempt of studying higher-$S$ spin chains, we  calculated the longitudinal spin structure factor for the model above with $D=0$ for   $S=1$ and $S=3/2$. The results for two representative  values of $\Delta$ are shown in Fig. \ref{fig:Sspins}. For both values of $S$ we notice a nearly dispersionless threshold in the spectral weight for $q\approx \pi$. This behavior is characteristic of finite-energy excitations with a large effective mass, which hinder the direct application of our   theory since they introduce a small band curvature energy scale.

\begin{figure}

\includegraphics[width=8cm]{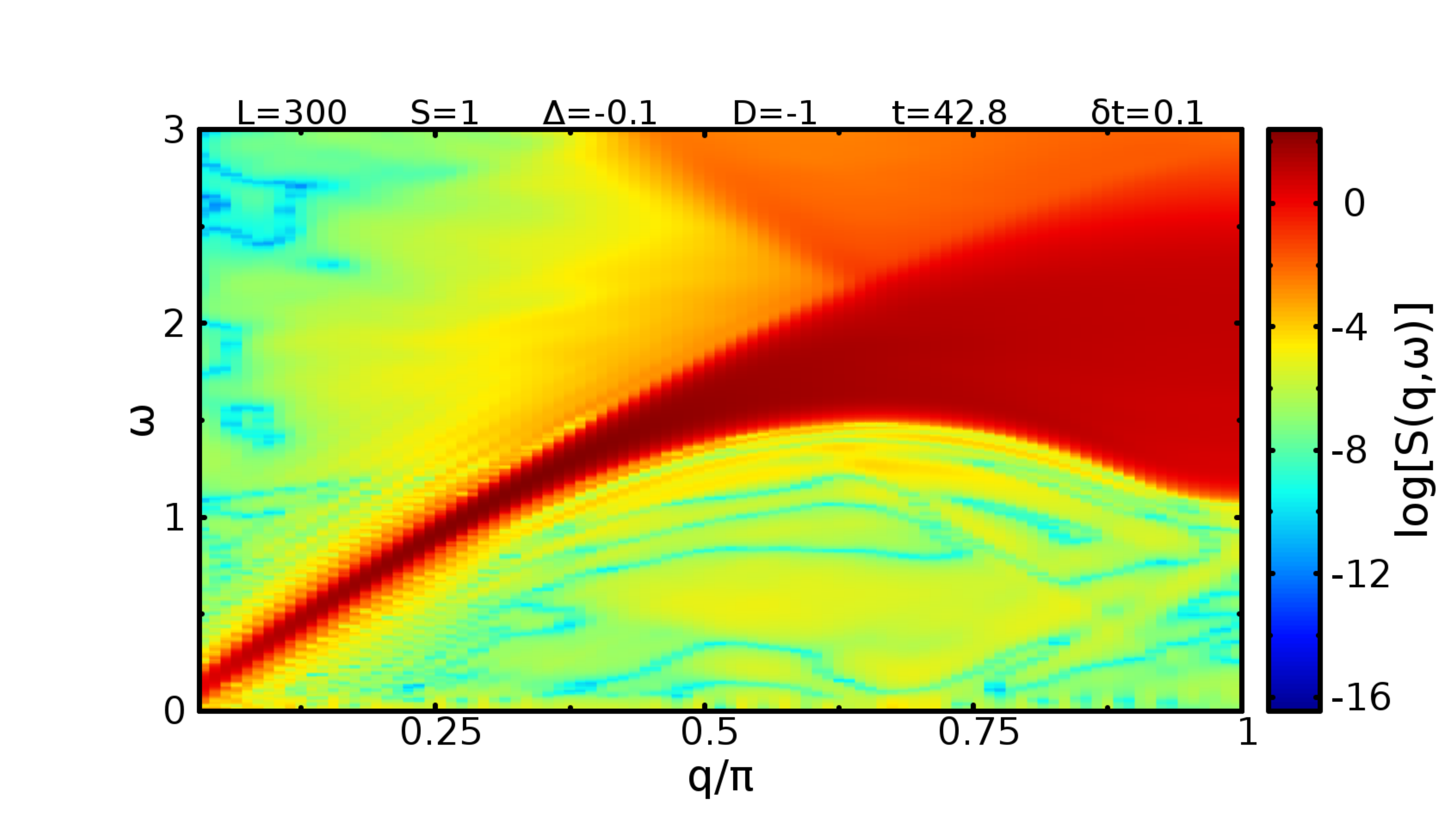}

\caption{(Color online)
The longitudinal spin structure factor of the spin-$1$ XXZ chain
with single-ion anisotropy for $\Delta=-0.1$ and $D=-1$. \label{fig:spin1finiteD}}
\end{figure}

Focusing on $S=1$ chains, we proceed by modifying the parameters in Eq. (\ref{higherS}) so as to look for a regime with a larger curvature of the spectrum near $q=\pi$. Remarkably, the gap in the spectrum of $S^\parallel(q\approx \pi,\omega)$  is consistent  with  the low-energy theory for critical spin-$1$ chains since the staggered part of the operator $S^z_j$ excites massive modes \cite{1985_Timonen_JPC_18,1986_Schulz_PRB_34}.  We consider the model with exchange anisotropy $\Delta=-0.1$ and easy-axis single-ion anisotropy $D=-1$, which lies  in the critical phase \cite{1993_Chen_PRB_67}. 
Fig. \ref{fig:spin1finiteD} shows  that in this case the lower threshold of $S^\parallel(q,\omega)$ has a smaller gap and larger band curvature at $q=\pi$. Note also that there is no evidence for bound states in the spectrum of Fig. \ref{fig:spin1finiteD}. 

%1 or 2 high-energy particles? regardless, exact power-law singularity at the threshold

\begin{figure}

\begin{centering}
\includegraphics[width=6.8cm]{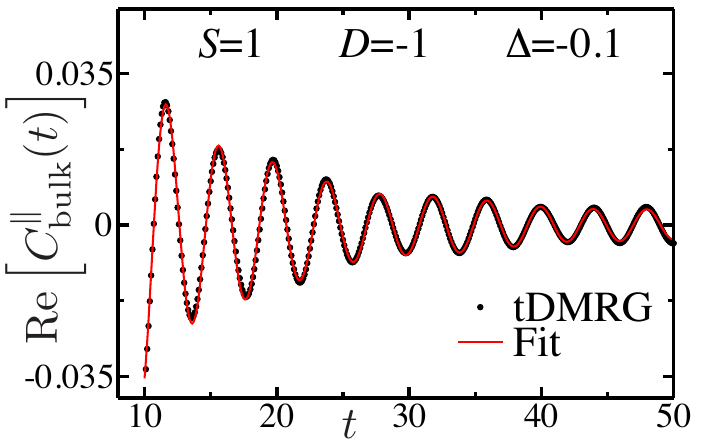}
\par\end{centering}

\caption{(Color online)
Real part of the  longitudinal spin autocorrelation $C^\parallel_{\text{bulk}}(t)$ vs. $t$ for the spin-$1$ chain with $\Delta=-0.1$, $D=-1$ and $L=300$. The data were obtained using $m=350$ and
$\delta t=0.1$. We fit the data to Eq. (\ref{twofreqfit}) and obtain
the frequencies $W_1=1.55$ and $W_2=1.11$ and exponents $\beta_1=1.57$
and $\beta_2=1.76$. \label{fig:bulkCtspin1}
}
\end{figure}

Next, we investigate the autocorrelation $C^\parallel(t,j)$ for the spin-1 chain  with $\Delta=-0.1$ and $D=-1$. As discussed in Section \ref{sec:integrability}, the sharp lower threshold of $S^\parallel(q,\omega)$ implies that   the bulk autocorrelation exhibits  power-law decay of its oscillating components. Note that this argument does not depend on details  of the mobile impurity model; the nonanalyticity in $C_{\text{bulk}}^\parallel(t)$ follows from integrating $S^\parallel(q,\omega)$ over momentum in the vicinity of the lower threshold. The frequencies of the oscillations can be read off from the spectrum of $S^\parallel(q,\omega)$ as the values of $\omega$ about which the lower threshold disperses parabolically. In the examples with spin-1/2 chains, there was only one such frequency corresponding to the band edge of single-hole excitations. By contrast, in Fig. \ref{fig:spin1finiteD} we observe   two frequencies that can be identified as ``edges'' of the support: $W_1\approx 1.5$ (at $q\approx 0.65\pi$) and  $W_2\approx 1.1$ (at $q\approx \pi$).  Thus, we have fitted the tDMRG data with the two-frequency  formula
\bea
\text{Re}\left[C_{\text{bulk}}^{\parallel}(t)\right]&=&\frac{B^{\parallel}_{0}}{t^{2}}+\frac{B^{\parallel}_{1}\cos(W_1t+\varphi_1)}{t^{\beta_1}}\nonumber\\&&+\frac{B^{\parallel}_{2}\cos(W_2t+\varphi_2)}{t^{\beta_2}}.\label{twofreqfit}
\eea
Note that in contrast with Eq. (\ref{eq:clongbulk}) here we include the nonoscillating term $\sim t^{-2}$, associated with the gapless $q=0$ mode, but omit the term $\sim t^{-2K}$ that in the spin-1/2 case stems from $q=\pi$ part of the operator $S_j^z$ in the LL theory. The result of the fit is shown in Fig. \ref{fig:bulkCtspin1}. Note that the frequencies obtained are consistent with the edges of the spectrum observed in Fig. \ref{fig:spin1finiteD}. 

\begin{figure}

\begin{centering}
\includegraphics[width=6.8cm]{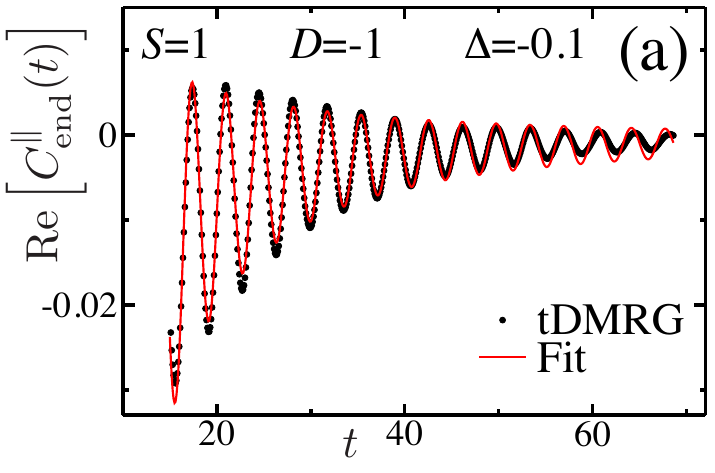}
\par\end{centering}

\begin{centering}
\includegraphics[width=6.8cm]{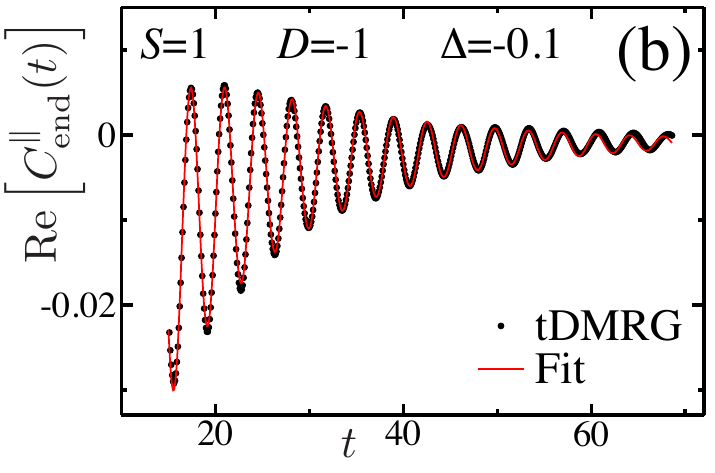}
\par\end{centering}

\caption{(Color online)
Real part of the  longitudinal spin autocorrelation $C^\parallel_{\text{end}}(t)$ vs. $t$ for the spin-$1$ chain with $\Delta=-0.1$, $D=-1$ and $L=300$. The symbols are the tDMRG results. The data were obtained using $m=350$ and
$\delta t=0.1$.(a) Fit to power-law decay in Eq. (\ref{f1fitpowerlaw}). (b) Fit to exponential decay in Eq. (\ref{f2fitexponential}).   \label{fig:endCtspin1}}
\end{figure}
 
Finally, we analyze the behavior of the boundary autocorrelation $C^\parallel_{\text{end}}(t)$ for the spin-$1$ chain   with $\Delta=-0.1$ and $D=-1$. For nonintegrable models our effective field theory predicts that boundary operators introduce a nonuniversal frequency  shift and  a decay rate for the high-energy mode. The numerical results  indicate that the data can be fitted with a single oscillating component.  We have fitted the tDMRG data for $C^\parallel_{\text{end}}(t)$ to two functions:\be
f_{1}(t)=\frac{A_1}{t^{2}}+\frac{A^{\text{pl}}_{2}\cos(W^\prime t+\varphi_1)}{t^{\beta}},\label{f1fitpowerlaw}\ee
versus\be
f_{2}(t)=\frac{A_1^\prime}{t^{2}}+A^{\text{exp}}_{2}\cos(W^\prime t+\varphi_2)e^{-\gamma t}.\label{f2fitexponential}\ee
For both fit functions we find   $W^\prime\approx 1.75$.  This frequency is clearly  different from the band edge frequencies $W_1$ and $W_2$ obtained from fitting  the bulk autocorrelation and  lies inside the continuum of $S^\parallel(q,\omega)$ (see Fig. \ref{fig:spin1finiteD}). This result is consistent with our prediction of a nonuniversal frequency shift for nonintegrable models. 
Moreover, we can see in  Fig. \ref{fig:endCtspin1}(a) that the best fit to Eq. (\ref{f1fitpowerlaw})   for $t>15$ overestimates the amplitude of the oscillations  at larger times $t\gtrsim 45$, suggesting that the decay is faster than power law. In fact, the fit to an exponential decay according to  Eq. (\ref{f2fitexponential})   with $\gamma\approx 0.059$ yields better agreement with the numerical data [see   Fig.  \ref{fig:endCtspin1}(b)]. Importantly, the fitted relaxation time $1/\gamma\approx 17$ is smaller than the time scales reached by the tDMRG.

\begin{figure}

\begin{centering}
\includegraphics[width=6.8cm]{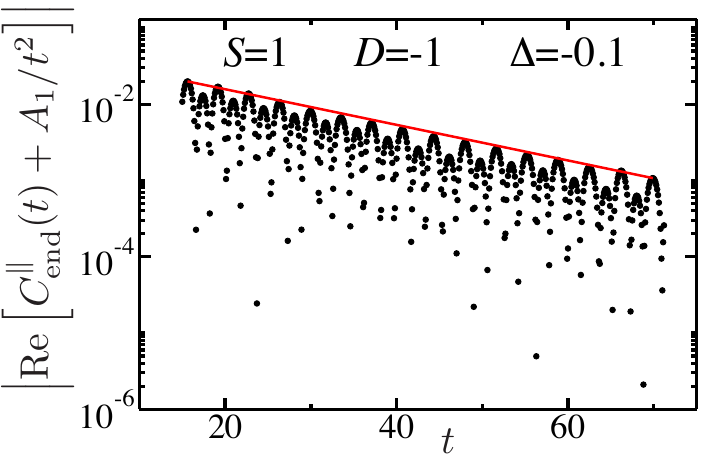}
\par\end{centering}

\caption{(Color online) 
Same  as Fig. \ref{fig:endCtspin1}, after subtracting the nonoscillating term $\sim 1/t^2$. The prefactor $A_1=2.233$ was obtained independently (see Appendix \ref{app:static}). The slope of the red line is $\approx -0.053$. \label{fig:loglinear}}
\end{figure}

In order to observe a clear signature of the exponential decay of $C^\parallel_{\text{end}}(t)$, it is convenient to subtract off the nonoscillating $t^{-2}$ term in the autocorrelation function. This subtraction is important because the difference between power-law and exponential decay of the oscillating component becomes more pronounced at longer times, after which an exponentially decaying term would become less significant  than the $1/t^2$ or subleading power-law terms. As explained in Appendix \ref{app:static}, we can fix the nonuniversal prefactor $A_1$ in Eq. (\ref{f1fitpowerlaw})  by relating it to the prefactor of the uniform term in the static correlation $\langle S_1^zS_j^z\rangle\sim 1/j^2$ for $j\gg 1$. The numerical result for the boundary autocorrelation after subtracting the nonoscillating term is shown in Fig. \ref{fig:loglinear}. It is clear  that the amplitude of the oscillations decays as a straight line on a log-linear scale.  This result indicates an exponential decay of the boundary autocorrelation in the nonintegrable model, in agreement with our prediction. 

\section{Conclusion\label{sec:conclusion}}

In conclusion, we have analyzed the effect of reflective boundary
conditions in one-dimensional quantum liquids on time-dependent
correlations. We have shown that one can generalize
the effective impurity model of a high-energy mode interacting with
the low-energy subband (nonlinear Luttinger liquid theory)
to capture the dominant contributions to late-time asymptotes of
autocorrelations and predict the exponents of associated power-law
singularities in the frequency domain. This was used to compute,
e.g., the autocorrelations in critical spin chains and the local
density of states at the band bottom in one-dimensional interacting
spinless fermions. The boundary exponents show a characteristic
doubling in their dependence on the phase shifts which implies
  relations between the bulk an boundary exponents depending
only on the Luttinger parameter but not on the
phase shifts. Generalizations of the method were used to derive
similar results for spinful models and different correlation
functions.

Our results apply, mutatis mutandis, to the class of integrable models,
but they need caution when applied to the nonintegrable case. While
the impurity mode is effectively protected in the bulk by momentum
conservation and power-law behavior of correlations is generic at zero
temperature, the breaking of translational invariance at the boundary
introduces the possibility of additional renormalization effects. We
have discussed two observable consequences: a shift in the impurity
energy leading to a shift in the oscillation frequency in the
autocorrelation, and the possibility of decay of the impurity leading to
exponential damping. These effects can be analysed within
the impurity model approach by studying boundary operators as
perturbations. Based on the Bethe ansatz solution for models with
reflective boundary conditions, we argue that integrable models should
be devoid of such effects and hence identical bulk and boundary
frequencies should be observed without exponential decay.  

We performed a time-dependent density matrix renormalization group
study of both integrable and non-integrable spin chains to verify our
predictions. For the integrable case, we studied the XXZ spin-1/2
chain and the numerically obtained correlations agree very well with
the effective field theory predictions. For the nonintegrable case we
looked at spin chains of higher spin $S>1/2$. We did find evidence
for a nonuniversal frequency shift in this case as well as an
exponential damping factor of the high-energy contribution to the
correlation. Detailed comparison with microscopic models
highlights the properties of the spectrum one should consider in
formulating the effective impurity model. First of all, one should
take into account all contributions from band minima as well as band
maxima. Complications may arrise when the spectrum features bound
states which are a priori not taken into account in the impurity
model and lead to additional oscillating contributions, but the impurity
model may in principle be adjusted to account for these. Bound-state lifetimes are subject to similar considerations
concerning the integrable versus nonintegrable case as the
high-energy impurity modes. A second complication comes when one of the
high-energy bands becomes nearly flat, resulting in a very large
time-scale before the asymptotic behavior of the correlation is
reached, which could possibly push it beyond the times for which
reliable numerical data can be obtained.

An experimental test of the oscillating, high-energy contribution to
correlations in real time would most likely involve the fabrication of
an effective spin model using cold atom systems, for which real-space and time-resolved
correlations can be imaged by many-body Ramsey interferometry
\cite{2013_Knap_PRL_111}.
To test our bulk versus boundary predictions one can resort to an
optical box-like potential
\cite{2013_Gaunt_PRL_110,2015_Chomaz_NATCOM_6} implementing the
appropriate boundary condition.

It would be interesting to extend our results to more general boundary
conditions. In particular, in the context of integrable models we may
distinguish between integrable and nonintegrable boundary conditions.
Moreover, one may differentiate between diagonal and non-diagonal
boundary conditions, the latter of which corresponds to boundary
conditions that do not conserve particle number in the fermionic
picture \cite{1988_Sklyanin_JPA_21,1994_Ghoshal_IJMPA_9,2015_Zhang_NPB_893}. The mobile impurity model, viewed as a boundary field theory,
in principle provides the flexibility to study all these situations
by choosing the appropriate boundary conditions as well as adding
boundary operators to account for possibly nontrivial boundary bound
states.

\begin{acknowledgments}
The work of ISE is part of the research programme of the Foundation for Fundamental Research on Matter (FOM), which is part of the Netherlands Organisation for Scientific Research (NWO). JCX and FBR acknowledge support by Brazilian agencies CNPq and FAPEMIG. RGP acknowledges support by CNPq.
\end{acknowledgments}

\appendix
\section{Boundary-bulk spin correlation \label{app:static}}
In this appendix, we relate the prefactors of the nonoscillating terms
of the time-dependent boundary autocorrelation and of the static spin correlation. 

Let us first consider the critical spin-1/2 XXZ chain with open boundary conditions. We are going to show
that the static spin correlation is given by\be
\langle S^z_1S^z_{j=x}\rangle \approx  -\frac{2\sqrt{K}A}{\pi ^2x^2}+\frac{B(-1)^x}{x^{1+K}},\label{answer}
\ee
where $K$ is the Luttinger parameter. 
The prefactor $A$ is nonuniversal and also appears in the time-dependent boundary autocorrelation\be
\langle S^z_1(t)S_1^z(0)\rangle\sim -\frac{4A^2}{\pi^2v^2t^2}+\text{oscillating terms}.\label{lowenergyautoSz}
\ee
Note that if  we determine the prefactor $A$ by fitting the numerical results for the static correlation to Eq.  (\ref{answer}), we can fix the prefactor   of the nonoscillating term in the time-dependent  boundary autocorrelation.

We start with the low-energy representation  for $S_j^z$ at the boundary:\bea
S_1^z&\sim &\Psi^\dagger(1)\Psi(1)\nonumber\\
&\sim &:\psi^\dagger_R(1)\psi^{\phantom\dagger}_R(1):+:\psi^\dagger_L(1)\psi^{\phantom\dagger}_L(1):\nonumber\\
&&+e^{i\pi}[\psi^\dagger_R(1)\psi^{\phantom\dagger}_L(1)+\text{h.c.}]\nonumber\\
&=&:\psi^\dagger_R(1)\psi^{\phantom\dagger}_R(1):+:\psi^\dagger_R(-1)\psi^{\phantom\dagger}_R(-1):\nonumber\\
&&+[\psi^\dagger_R(1)\psi^{\phantom\dagger}_R(-1)+\text{h.c.}]\nonumber\\
&\sim&4:\psi^\dagger_R(0)\psi^{\phantom\dagger}_R(0):\nonumber\\
&\sim&-\frac{4}{\sqrt{2\pi}}\partial_x\phi_R(0).\label{phi0}
\eea
Next, we need to perform the Bogoliubov transformation:\be
\phi_R(x)=\frac{K^\frac12+K^\frac12}{2}\varphi_R(x)-\frac{K^\frac12-K^\frac12}{2}\varphi_R(-x).\label{bogol}
\ee
In the interacting  case, the boundary operator has a nonuniform prefactor because the expression in  Eq. (\ref{phi0})  mixes  the staggered part of the density operator $\psi^\dagger_R\psi^{\phantom\dagger}_L+\text{h.c.}$  (which has a nonuniversal prefactor when bosonized  in the interacting case) with the uniform part  $\psi^\dagger_R\psi^{\phantom\dagger}_R+\psi^\dagger_L\psi^{\phantom\dagger}_L$ (which does have a universal prefactor). For this reason,  in the general case we must write\be
S_1^z\sim -\frac{4A}{\sqrt{2\pi}}\partial_x\varphi_R(0),\label{boundaryoperator}
\ee
where $A=1$ for  free fermions, but $A$ is nonuniversal  in the interacting case.  Using Eq. (\ref{boundaryoperator}) together with the bosonic propagator,\bea
\langle \partial_x\varphi_R(x,t)\partial_x\varphi_R(0,0) \rangle=-\frac{1}{2\pi(x-vt)^2},
\eea
leads to the result in Eq. (\ref{lowenergyautoSz}).

The spin operator in the bulk is given by \bea
S_{j=x}^z&\sim &\Psi^\dagger(x)\Psi(x)\nonumber\\
&\sim &\psi^\dagger_R(x)\psi^{\phantom\dagger}_R(x)+\psi^\dagger_L(x)\psi^{\phantom\dagger}_L(x)\nonumber\\
&&+(-1)^x[\psi^\dagger_R(x)\psi^{\phantom\dagger}_L(x)+\text{h.c.}]\nonumber\\
&\sim&\sqrt{\frac{K}{2\pi}}[\partial_x\varphi_L(x)-\partial_x\varphi_R(x)]\nonumber\\
&&+\frac{(-1)^x}{2\pi\eta} \left[e^{i\sqrt{2\pi K}[\varphi_R(x)-\varphi_L(x)]}+\text{h.c.}\right].
\eea
Using the folding trick with \be
\partial_x\varphi_L(x)=-\partial_x\varphi_R(-x),\label{folddxphi}
\ee
we obtain  \bea
\hspace*{-0.4cm}S_{j}^z&\sim&-\sqrt{\frac{K}{2\pi}}[\partial_x\varphi_R(x)+\partial_x\varphi_R(-x)]\nonumber\\
&&+B^\prime(-1)^x \left[e^{i\sqrt{2\pi K}[\varphi_R(-x)-\varphi_R(x)]}+\text{h.c.}\right],\label{Szbulk}
\eea
where $B^\prime$ is nonuniversal.

Let us first focus on the uniform part in Eq. (\ref{Szbulk}). The corresponding term in the static correlation is\bea
\langle S_1^zS_j^z\rangle&\sim& \frac{2\sqrt{K}A}{\pi}\left[\langle \partial_x\varphi_R(0)\partial_x\varphi_R(x)\rangle+(x\to-x)\right]\nonumber\\
&=&-\frac{2\sqrt{K}A}{\pi^2x^2},
\eea
which is the first term on the rhs of Eq. (\ref{answer}).

Now consider the staggered part of the operator in Eq. (\ref{Szbulk}). Since this term has a nonuniversal prefactor which is  independent of $A$, we shall  focus on deriving  the exponent of the large-distance decay. The staggered term in the correlation is   \bea
\langle S_1^zS_j^z\rangle&\sim& (-1)^x\langle \partial_x\varphi_R(0)e^{i\sqrt{2\pi K}\varphi_R(-x)}e^{-i\sqrt{2\pi K}\varphi_R(x)}\rangle.\nonumber\\
&&
\eea
This is a three-point function involving three primary fields. We   use the operator product expansion:\bea
&&:\partial_x\varphi_R(0):\,:e^{i\sqrt{2\pi K}\varphi_R(-x)}:\nonumber\\
&=&\sum_{n=0}^{\infty}\frac{(i\sqrt{2\pi K})^n}{n!}:\partial_x\varphi_R(0):\,:[\varphi_R(-x)]^n:\nonumber\\
&\sim&\sum_{n=1}^{\infty}\frac{(i\sqrt{2\pi K})^n}{(n-1)!}\langle \partial_x\varphi_R(0)\varphi_R(-x)\rangle \,:[\varphi_R(-x)]^{n-1}:\nonumber\\
&=&i\sqrt{2\pi K}\langle \partial_x\varphi_R(0)\varphi_R(-x)\rangle :e^{i\sqrt{2\pi K}\varphi_R(-x)}:\nonumber\\
&=&\frac{i\sqrt{K}}{\sqrt{2\pi} x}\,:e^{i\sqrt{2\pi K}\varphi_R(-x)}:.
\eea
Thus, in the three-point function we obtain\bea
&&\langle \partial_x\varphi_R(0)e^{i\sqrt{2\pi K}\varphi_R(-x)}e^{-i\sqrt{2\pi K}\varphi_R(x)}\rangle\nonumber\\
&\sim& \frac1x\langle e^{i\sqrt{2\pi K}\varphi_R(-x)}e^{-i\sqrt{2\pi K}\varphi_R(x)}\rangle\nonumber\\
&\sim&\frac{1}{x}\frac1{(2x)^K}.
\eea
It follows that the staggered term in the spin correlation behaves as\be
\langle S_1^zS_j^z\rangle\sim \frac{(-1)^x}{x^{1+K}}, 
\ee
which is the second term   in Eq. (\ref{answer}).

For the spin-1 chain the uniform part of the spin operator in the bulk
becomes\be
S_{j}^z\sim-\sqrt{\frac{K}{\pi}}[\partial_x\varphi_R(x)+\partial_x\varphi_R(-x)].\label{Szspin1}
\ee Note the extra factor of $\sqrt{2}$ in comparison with
Eq. (\ref{Szbulk}), which comes from combining the densities of two
spinless fermions \cite{1986_Schulz_PRB_34} (more generally, this
procedure introduces a factor of $\sqrt{2S}$ for the spin-$S$
operator). The Luttinger parameter in Eq. (\ref{Szspin1}) is defined
such that the Kosterlitz-Thouless transition to the gapped Haldane
phase happens at $K=1$ and $K>1$ in the critical phase
\cite{1986_Schulz_PRB_34}.  Moreover, for $S=1$ the staggered part of
$S^z_j$ couples to gapped modes (recall the spectrum is gapped at
$k=\pi$). As a result, the staggered term in the static correlation
decays exponentially with the distance from the boundary. The results
for the autocorrelation and static correlation for $S=1$ are\bea
\langle S^z_1(t)S_1^z(0)\rangle&\approx&
-\frac{4C^2}{\pi^2v^2t^2},\\ \langle S^z_1S^z_{j=x}\rangle &\approx&
-\frac{2\sqrt{2K}C}{\pi ^2x^2},\label{eqA17}\eea where the coefficient
$C$ is nonuniversal. The LL parameter $K$ and the spin velocity $v$
can be determined independently by analyzing the finite-size
corrections of the lower energy states together with the machinery of
the conformal field theory \cite{DiFrancescoBOOK}, see for example
Ref. \cite{2010_Xavier_PRB_81}. We found   for the spin-1 chain
with $\Delta=-0.1$ and $D=-1$ the following values: $K=1.285$ and
$v=1.211$. Using these values and fitting the DMRG data of the static
correlations to Eq. (\ref{eqA17}), we found that $C=2.8423$.

\bibliography{boundaryBIB}% Produces the bibliography via BibTeX.

\end{document}